\documentclass[10pt,journal,compsoc]{IEEEtran}
\usepackage{subfigure}
\usepackage{multirow}
\usepackage{booktabs}
\usepackage{paralist}
\usepackage{amsmath}
\usepackage{array}
\usepackage{lscape}

\newcommand{\PreserveBackslash}[1]{\let\temp=\\#1\let\\=\temp}
\newcolumntype{C}[1]{>{\PreserveBackslash\centering}p{#1}}
\newcolumntype{R}[1]{>{\PreserveBackslash\raggedleft}p{#1}}
\newcolumntype{L}[1]{>{\PreserveBackslash\raggedright}p{#1}}

\newcommand{\tabincell}[2]{\begin{tabular}{@{}#1@{}}#2\end{tabular}}

\ifCLASSOPTIONcompsoc
  \usepackage[nocompress]{cite}
\else
  \usepackage{cite}
\fi

\ifCLASSINFOpdf
   \usepackage[pdftex]{graphicx}
   \DeclareGraphicsExtensions{.pdf,.jpeg,.png}
\else
   \usepackage[dvips]{graphicx}
   \DeclareGraphicsExtensions{.eps}
\fi

\begin{document}
%
\title{Sound Synthesis, Propagation, and Rendering: A Survey}

\author{Shiguang~Liu,~\IEEEmembership{Senior Member,~IEEE,} 
        Dinesh Manocha,~\IEEEmembership{Fellow,~IEEE} 
\IEEEcompsocitemizethanks{\IEEEcompsocthanksitem S. Liu was with the School of Computer Science and Technology, College of Intelligence and Computing, Tianjin University, Tianjin 300350,
P.R. China. E-mail: lsg@tju.edu.cn \protect\\
D. Manocha was with Department of Computer Science, University of Maryland at College Park, College Park, MD 20742 USA, and also with the Department of Electrical and Computer Engineering, University of Maryland at College Park, College Park, MD 20742 USA. 
E-mail: dm@cs.umd.edu

}
}


\IEEEtitleabstractindextext{%
\begin{abstract}
Sound, as a crucial sensory channel, plays a vital role in improving the reality and immersiveness of a virtual environment, following only vision in importance. Sound can provide important clues such as sound directionality and spatial size. This paper gives a broad overview of research works on sound simulation in virtual reality, games, etc. We first survey various sound synthesis methods, including harmonic synthesis, texture synthesis, spectral analysis, and physics-based synthesis. Then, we summarize popular sound propagation techniques, namely wave-based methods, geometric-based methods, and hybrid methods. Next, the sound rendering methods are reviewed. We also highlight some recent methods that use machine learning techniques for synthesis, propagation, and some inverse problems.  
\end{abstract}

}

\maketitle

\IEEEdisplaynontitleabstractindextext

%
\IEEEpeerreviewmaketitle

\IEEEraisesectionheading{\section{Introduction}\label{sec:introduction}}

\IEEEPARstart{S}{ound} simulation deals with synthesis, propagation and rendering of audio effects. It is well known that modeling the sound effects and spatialized audio rendering can significantly enhance the sense of realism and presence in virtual environments. Moreover, physically-based sound generation methods can be used to synthesize realistic and synchronized sounds for physically-based animation and virtual environments. Most of the work in computer graphics, games, computer-aided design (CAD), virtual environments, etc. has focused on visual rendering and generating photorealistic images. These visual rendering methods are commonly used in commercial and research systems. Over the last few decades, research in sound simulation has received significant attention in terms of sound synthesis, sound propagation, and sound rendering (Fig. 1). In this survey, we give a brief  overview of sound simulation methods


\begin{figure}[!h]
\centering
\includegraphics[width=3.5in]{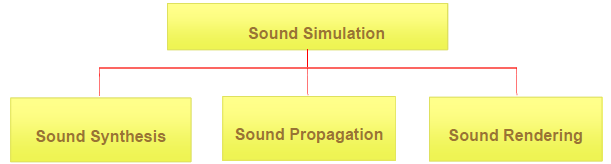}
\caption{Overview of the sound simulation research.}
\label{fig0}
\end{figure}

\begin{figure*}[!t]
\centering
\includegraphics[width=5.5in]{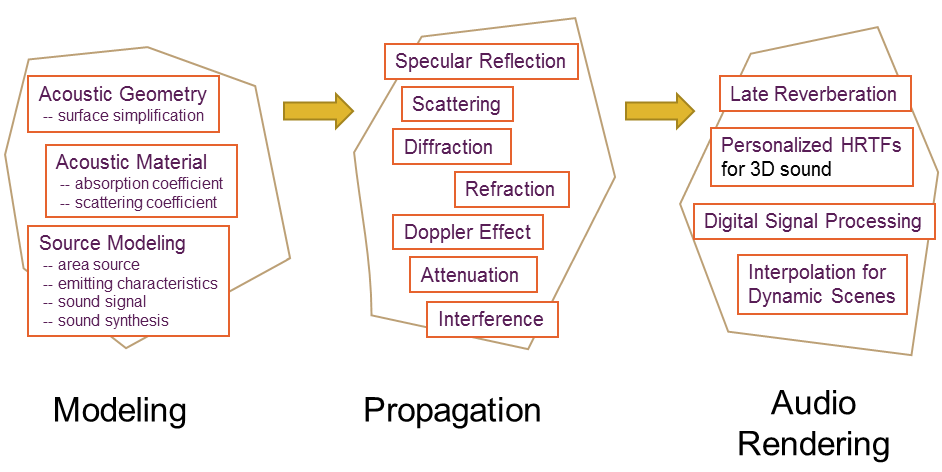}
\caption{General sound simulation pipeline used in games, VR, or multimedia applications. There is substantial work within each area and we give an overview of many techniques.}
\label{fig00}
\end{figure*}

A sound wave is produced by the vibration of an object as it spreads through a medium (air, solid, or liquid) or by pressure variations in the gas itself such as the spoken voice, and this wave can be perceived by human ears. Sound travels in the form of waves. Objects that initially vibrate are sound sources. This paper focus on environmental sound processing techniques and does not pay attention to speech processing methods or musical instrument sound synthesis methods. As shown in Figure \ref{fig0}, sound simulation research in computer graphics and virtual reality has mainly focused on synthesis of sound sources, sound propagation in the environment, and sound rendering. Figure \ref{fig00} highlights the sound simulation pipeline.
 
In earlier works, sound was modeled by recording, which is easy to implement. However, it can only generate a fixed sound effect, instead of being able to distinguish the source of the sound, position of the listener, or the scale of the space. In such cases, when the visual rendering changes, the sound source in the recording remains fixed, which greatly affects the sense of presence in VR (Virtual Reality) systems. In addition, with recordings, it is not possible to fully simulate the sound generated for each scene. For example, it is difficult to simulate the sound effects corresponding to complex collisions among objects with varying materials such as wood, plastic, metal, fluids, etc. Moreover, it is not easy to deal with sound propagation effects in a virtual scene with only using such recordings. Lastly, manual synchronization between visual animation and sound frame by frame also represents a significant challenge for the recording method. To this end, researchers began to develop various sound synthesis and sound propagation methods. 

As far as we know, in 1992 Takala and Hahn \cite{Tak80} proposed one of the first sound rendering pipeline, which aims to produce the composite soundtrack from component sound objects. Their method can be regarded as a sonic extension of the known image rendering 
pipeline because it views sound as a one-dimensional signal regularly sampled in time. The proposed pipeline includes generation, representation, and synchronization of sounds for a behavioral or physics-based animation. Later, researchers developed advanced sound synthesis methods including harmonic synthesis \cite{Gor00}, texture synthesis \cite{Dub02}, spectral analysis \cite{Ser90}, and physics-based synthesis \cite{Doe96}. Various types of sounds, such as rigid-bodies, liquids, fire, fractures, and aerodynamic phenomena can be realistically synthesized. 

Sound propagation modeling is usually described using the wave equation, which also has time domain and frequency domain formulations. The propagation of sound in a medium can be described by the following acoustic wave function \cite{Pie81}
\begin{equation}
\nabla^2p - \frac{1}{c^2} \frac{\partial^2 p}{\partial t^2} = f,
\end{equation} 
where $p=p(x, t)$ denotes the pressure at location $x$ and time $t$, $f(x, t)$ represents a force function generated by the sound source, $c$ is the speed of sound, and $c \approx 343$ m/s. In the frequency domain, sound propagation can be formulated as a boundary value problem by the Helmholtz equation, written as
\begin{equation}
\nabla^2p + \frac{\omega^2 }{c^2} p = 0, x \in \Omega
\end{equation} 
where $p=p(x, \omega)$ is the complex-valued pressure field, $\omega$ represents the angular frequency, $\Omega$ is the propagation domain, and $\nabla^2$ denotes the Laplacian operator. The Helmholtz equation is usually used when the sound source is time harmonic (at one frequency). In some cases, a frequency sweep is performed (first generate a Helmholtz solution at multiple frequencies, then do an inverse Fourier transform). Wave based methods can accurately solve this equation. In order to simulate sound propagation in an efficient manner, under a high-frequency assumption, geometry-based methods model the acoustic effects in an environment based on the geometric construction of rays via the ray theory in geometry optics. There are also research works that take advantage of both wave-based methods and geometry methods to reach a better balance in terms of quality and efficiency \cite{Fun03}. By setting the initial value and boundary conditions to solve the Helmholtz equation, the wave-based method can capture the diffraction, interference, scattering, acoustic focusing, and high-order wave effects of sound waves to obtain accurate analytical solutions. 

Sound rendering and spatialization are crucial for producing realistic sound effects in virtual environments \cite{Tsi04}. The computational bottlenecks of sound rendering can be classified into two aspects, the cost of spatialization, and per sound source cost. The former includes using the Head Related Transfer Function (HRTF) for binaural rendering, processing a large number of speakers for a WFS (Wave Field Synthesis) system, etc. The latter includes the calculation about the Doppler effect, sound delay, reverberation, etc. 

\begin{table}[htb]
\centering

\caption{Prior Surveys on Sound Simulation}
\label{table0}
\setlength{\tabcolsep}{0.16in}
\begin{tabular}{cc}
\hline
\specialrule{0em}{0.5pt}{0.5pt}
Surveys                          & Topics           \\  
\specialrule{0em}{0.5pt}{0.5pt}
\hline 
\specialrule{0em}{0.5pt}{0.5pt}
Kleiner et al. 1993 \cite{Kleiner93}                & \tabincell{c} {Sound auralization \\ Sound propagation 
}\\
\specialrule{0em}{0.5pt}{0.5pt}
\hline
\specialrule{0em}{0.5pt}{0.5pt}
Cook 2002 \cite{Cook2002}  & Sound synthesis \\
\specialrule{0em}{0.5pt}{0.5pt}
\hline
Lokki et al. 2002 \cite{Lokki2002} & \tabincell{c}{Sound rendering \\ Room acoustic simulation \\ 
Auralization} \\
\specialrule{0em}{0.5pt}{0.5pt}
\hline
Funkhouser et al. 2003 \cite{Fun03} & \tabincell{c} {Sound propagation \\  Room acoustic simulation \\ 
Auralization} \\
\specialrule{0em}{0.5pt}{0.5pt}
\hline
Savioja et al. 2010 \cite{Sav10} &  \tabincell{c} {Room acoustic modeling \\ Auralization} \\
\specialrule{0em}{0.5pt}{0.5pt}
\hline
\specialrule{0em}{0.5pt}{0.5pt}
V$\ddot{a}$lim$\ddot{a}$ki et al. 2012 \cite{Par12} & Artificial reverberation \\
\specialrule{0em}{0.5pt}{0.5pt}
\hline
Roginska 2017 \cite {Rog17} & \tabincell{c} { Sound rendering\\ Spatial sound} \\
\specialrule{0em}{0.5pt}{0.5pt}
\hline
Serafin et al. 2018 \cite{Serafin2018} & \tabincell{c}{Sound rendering \\ Sound propagation} \\
\specialrule{0em}{0.5pt}{0.5pt}
\hline
Savioja and Xiang 2019 \cite{Savioja2019} & \tabincell{c} {Room acoustic modeling\\ Auralization } \\
\specialrule{0em}{0.5pt}{0.5pt}
\hline
Serafin et al. 2020 \cite{Serafin2020} & \tabincell{c}{Sound rendering \\ Sound propagation}   \\
\specialrule{0em}{0.5pt}{0.5pt}
\hline
Ours & \tabincell{c} {Sound synthesis \\ Sound propagation \\Sound rendering} \\

\hline

\end{tabular}
\end{table}

As shown in Table \ref{table0}, in contrast to specific surveys on sound simulation \cite{Kleiner93}\cite{Cook2002}\cite{Lokki2002}\cite{Fun03}\cite{Sav10}\cite{Par12}\cite{Sav15}\cite {Rog17}\cite{Serafin2018}\cite{Savioja2019}\cite{Serafin2020}, we present a broader review on sound simulation research from the perspective of sound synthesis (Sec. 2), sound propagation (Sec. 3), and sound rendering (Sec. 4). We also cover many recent developments and techniques that tend to combine synthesis and propagation, propagation and rendering, etc. Moreover, we summarize the latest work on sound simulation using machine learning methods (Sec. 5). Finally, we highlight some new and future directions in this field, including inverse problems (Sec. 6). 

\section{Sound synthesis}

\subsection{Harmonic synthesis methods}

This type of method uses existing sample data or experimental results and a set of basic functions to generate sound. It does not need on-line calculation and only needs to choose the appropriate interpolation and fusion method. Therefore, the calculation speed is very fast, and this method was very popular in the early stages of sound synthesis research. 

Gordon \cite{Gor00} pointed out that any sound waveform can be represented as a set of sine waves. Given a sound with a harmonic structure, each of those sine waves has a frequency that is an integer multiple of the fundamental frequency. According to the Fourier theory, this method uses harmonic components of different frequencies and amplitudes to synthesize the sounds of various musical instruments, also called additive synthesis.

\subsection{Sound texture synthesis methods}

In these methods, sound textures of any desired duration are synthesized by a given sampled texture.

Granular synthesis is a popular texture synthesis method \cite{Bar88}\cite{Roa04}. A grain is a small piece of sonic data, with a duration generally between 10 to 50 ms. A piece of new sound can be synthesized by redistributing and reorganizing the tiny grains. In contrast, the traditional cutting and splicing method is based on a given sound example. This method can provide a powerful tool to create unheard sounds and textures, and it is often exploited by sound designers. 

The wavelet tree learning method \cite{Dub02} regarded sound as a series of short, distinct bursts of energy. Given an input sound texture, a new sound can be synthesized by repeating structural elements (i.e., sound grains, represented as a tree) subject to some randomness in terms of time and ordering but maintaining certain essential temporal coherence and across-scale localization. This method can reproduce random, periodic sound signals, but it is difficult to control the synthesized sound effect through user input. 

Parker et al. \cite{Par04} processed the sound samples by cutting and splicing, thus generating a new sound texture. Schwarz et al. \cite{Sch08} proposed a descriptor based sound texture sampling method, which cuts sound into many small segments, and then selected the loudness, periodicity, and spectral characteristics of each segment of sound as descriptors. By controlling the value of one or more descriptors, it can generate sound of any length. They first successfully synthesized the sound of rain. Later, they \cite{Sch14} synthesized the sound of wind. This kind of method can produce new sound textures with good quality for all the objects studied, but the process of texture generation is not combined with specific animation. Therefore, this kind of work can be applied to scene sound generation as a sample expansion method in the preparation stage of a sound bank.

Cardle et al. \cite{Car03} proposed a new automatic, motion driven sound synthesis technique. The method divides the motion signal for training and maps the sound signal to the input motion signal (such as a moving car). This method can synthesize the sound of a two-dimensional fluid, but it is difficult to extend it to three-dimensional cases, which greatly limits its practical application. Strobl et al. \cite{Str06} gave a good overview and comparative analysis of the sound signal generation technology based on texture synthesis. Picard et al. \cite{Pic06} proposed a sound particle mixing method for generating specific sound effects. This kind of sample-based method has a simple process with high synthesis efficiency, but it needs more parameter control, and the acquisition of parameters often needs manual intervention. Sterling and Lin \cite{Ste16} further improved the method by introducing a vector graph \cite{Ste15} and high-order damping attenuation.

Recently, Zhang et al. \cite{Zhang19} proposed the acoustic texture rendering method for extended stochastic sources, e.g., falling rain or a flowing waterway, through the event loudness density (ELD), which relates the rapidity of received events to their loudness. By formulating the ELD as a function of listener location, realistic spatial variation in texture can finally be achieved.

\subsection{Spectral analysis methods}

Serra and Smith \cite{Ser90} proposed a spectral modeling synthesis method. In this method, the sound spectrum is divided into two parts, the deterministic spectrum and the stochastic spectrum. By combining these two parts of a spectrum, the resulting sound can be derived. This method can be used to synthesize noise such as the roar, hiss, and burst of a fireplace animation. Marelli et al. \cite{Mar10} extended the method and proposed a sound generation model based on narrow band spectrum to synthesize the silky sound of flame and the crackle sound of a slight crack in a furnace. McDermott et al. \cite{Mcd09} synthesized flame sounds by spectral noise. The results show that the low-frequency sound information of flames can be well expressed, and the lost high-frequency sound information can be compensated for by high order statistics. Because this kind of method is based on the sound spectrum rather than the fluid motion simulation data, it is difficult to maintain the synchronization with the fluid motion state.

\subsection{Physics-based synthesis methods}
Although the above non-physical methods can produce perceptually plausible sound, most of them need laborious manual control. Moreover, those methods make it difficult to synchronize sound with visual animation and therefore incur sensory experiences for users. 

In contrast, physically-based methods generally produce sound by first extracting animation information such as the position information and motion state of the object, then using the corresponding acoustic equation to calculate the sound pressure so that the generated sound can be better synchronized with visual rendering without additional manual editing. Currently, physically-based methods are widely used for synthesizing rigid-body sounds, flame sounds, liquid sounds, aerodynamic sounds, etc.

\begin{figure}[!t]
\centering
\includegraphics[width=3.5in]{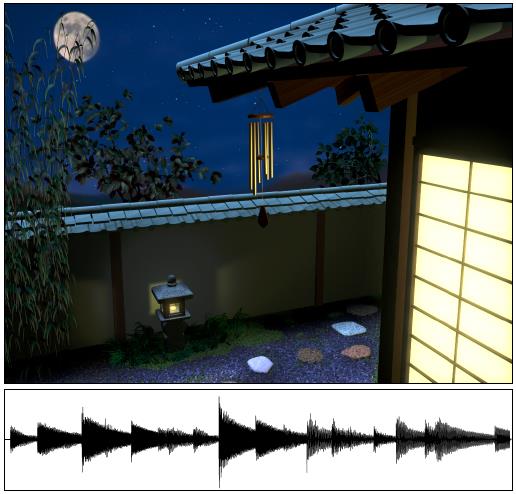}
\caption{A set of simulated wind chimes in a synthetic environment \cite{Obr02}.}
\label{fig1}
\end{figure}

\subsubsection{Rigid-body sound}  Rigid-body sound synthesis has been of great interest for general aural content. The \textsl{modal sound synthesis method}, based on modal analysis, can effectively generate the sound of rigid-body contact. When an object vibrates, it will radiate sound pressure waves into the surrounding air to meet the listener. The sound waves are produced by air pressure fluctuations caused by object vibrations. The modal sound synthesis decouples the object vibration into independent modes and evaluates the loudness of each mode. The sound of a vibration mode at position $x$ and time $t$ can be expressed by the weighted linear superposition of the modal oscillations, written as 
\begin{equation}
p(x,t)=\sum_j a_j(x)q_j(t)
\end{equation}
where $a_j(x)$ is the acoustic transfer function.

Van Den Doel and Pai \cite{Doe96} proposed an analytical model of linear modal vibration to generate contact sound in interactive virtual environments, which made modal sound synthesis popular. Modal sound models can be easily estimated from recordings and measurements \cite{Doe01}, the linear modal analysis using a numerical technique \cite{Obr02}, or the example-guided physically-based synthesis method \cite{Ren13}. 

\begin{figure}[!t]
\centering
\includegraphics[width=3.5in]{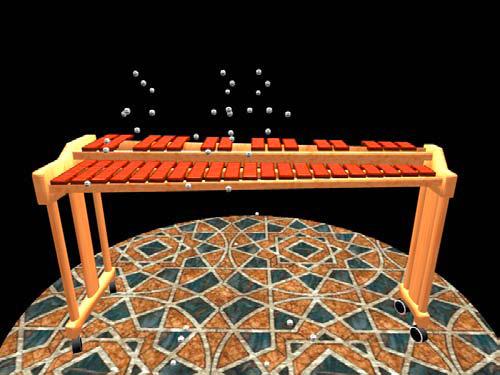}
\caption{Numerous dice fall on a three-octave xylophone in close succession, playing out the song "The Entertainer" \cite{Rag06}.}
\label{fig2}
\end{figure}

Van Den Doel et al. \cite{Doe01} realized the importance of sampling the micro impact force at the (near) audio rate, and clearly distinguished the so-called "dynamic force" of sampling at the graphics rendering rate. O'Brien et al. \cite{Obr01} proposed a method for approximating sounds generated by the motion of solid objects. They used physics-based animation to model the motion of solid objects and analyzed their surfaces using a deformable body simulator to judge if sound pressure waves may be induced in the surrounding air. Different from heuristic methods (e.g., harmonic synthesis methods) that are specific to particular objects, this method tackled the problem of automatic generation of sound synthesis for general deformable rigid-bodies, which is an important step of sound synthesis. O'Brien et al. \cite{Obr02} used tetrahedral finite element simulation to synthesize the modal sound of rigid-bodies under the contact force (see Figure \ref{fig1} for the result of a set of simulated wind chimes in a synthetic environment). However, the rigid-body model can not capture deformable contact events, and the computational complexity is high when large-scale rigid bodies are in contact collision at the same time. Linear modal synthesis methods suffer from a lack of automatic evaluation of satisfactory material parameters. Ren et al. \cite{Ren13} proposed an example-guided physically-based method by estimating salient sonic features from given sound examples. This method, grounded on auditory perception~\cite{6479182}, can produce realistic-sounding results synchronized with different materials and dynamics. Sterling et al.~\cite{Ste19} introduced an improved method using probabilistic damping model for extracting material parameters from recorded audio. In order to improve the efficiency of modal sound synthesis, some simplified (modal) sound models have been proposed. For example, Hahn et al. \cite{Hah14} introduced the timbre tree into the parametric sound model. Other extensions of modal contact sound synthesis include the trade-off between speed and accuracy of sound in interactive applications \cite{Doe04}\cite{Rag06}\cite{Bon08}\cite{Ren10}. Moreover, acoustic transmission model  is used to improve the authenticity of sound in three-dimensional space \cite{Jam06} and Raghuvanshi et al. \cite{Rag06} proposed a sound quality scaling scheme based on the priority level. This method makes use of the auditory perception effect of the human ear to satisfy the strict time limits in interactive applications, while ensuring the sound quality is maintained as much as possible. The authors designed a three-octave xylophone, as shown in Figure~\ref{fig2}, where many dice fall onto the keys to produce the corresponding musical notes. However, this method only considers the transient sound produced by the collision of rigid bodies, ignoring the sound produced by the breaking of rigid bodies and the rolling of objects.

Van Den Doel et al. \cite{Doe01} designed a synthesis method for contact sounds produced by collision, sliding, and rolling between rigid bodies. They take the contact force generated or customized by simulation as input, and the user selects the corresponding modal model to calculate different sound effects. Ren et al. \cite{Ren10} proposed a new contact processing model for contact sound simulation. The model uses a three-layer surface representation to describe the physical surface, which can generate more accurate and delicate contact sound. Zheng and James \cite{Zheng11} proposed a new method for high-quality modal contact sound synthesis via frictional multibody contact formulation. They modified the Staggered Projections solver so that noise artifacts associated with spatial and temporal contact-force fluctuations can be alleviated. 
Interactive sound synthesis techniques have also been incorporated into design of virtual instrument systems on multi-touch tabletop systems~\cite{Ren2012a,Ren2012b}, consumer mobile devices~\cite{Ren15}, and use of 3D printing 
technologies~\cite{Bha15}. These systems allows users to design and customize musical instruments and perform musical effects in real time. 

Physically-based sound simulation methods often require users to manually adjust some parameters to achieve satisfactory sound effects, a process that not only takes time, but also causes a lot of repeated calculations. Additionally, with current linear modal sound models, when synthesizing different sound characteristics by tweaking modal parameters, one needs to change the modal sound model completely through expensive re-computation. This method greatly eases the parameter tuning without expensive re-computation. The sounding results are comparable to those produced by expensive re-computation. Li et al. \cite{Li15} proposed an interactive sound transmission approximation method, that effectively avoided the problems of traditional methods described above. Although modal sound models have achieved great success, they still suffer from high memory consumption. To this end, Langlois et al. \cite{Lan14} proposed a method for compressing modal sound models, that can achieve compression ratios in the hundreds while preserving comparable audible sound results. 

\begin{figure*}[!t]
\centering
\includegraphics[width=7.0in]{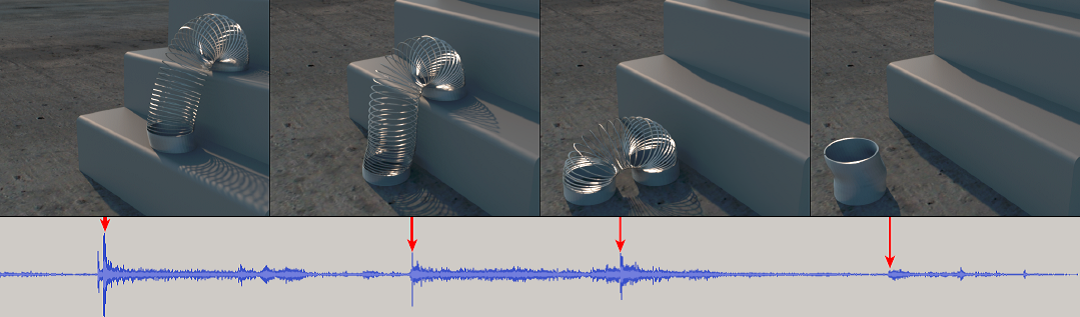}
\caption{Sound synthesis of a spring toy slinking down the stairs \cite{Sch17}.}
\label{fig3}
\end{figure*}

Due to large displacement, nonlinear vibrations, and the geometrically singular nature of rods, the above methods may fail when synthesizing the sound of elastic rods, which are an important animation primitive in CG. Schweickart et al. \cite{Sch17} synthesized the sound for elastic rod animation by leveraging the Kirchhoff theory, which was used to simplify the representation of rod dynamics. Moreover, a generalized dipole model was designed to compute the acoustic radiation. Figure \ref{fig3} shows a simulation result for a spring toy slinking down the stairs. In this case, 3888 control points are sampled for this discrete elastic rod from tens of thousands of complex collision events in each timestep.

In modal sound synthesis for rigid-body objects, it is necessary to evaluate material parameters that can approximate real ones. Sterling et al. \cite{Ste19} proposed estimating the damping parameters of materials from recorded impact sounds, greatly alleviating the human effort.

\textbf{Precomputed acoustic transfer methods.} When sound radiates from vibrating objects, perceptible diffraction and interreflection occur. It is challenging to simulate these phenomena at audio rates. James et al. \cite{Jam06} proposed Precomputed Acoustic Transfer (PAT) for real time synthesis of sound radiation from rigid objects. This method consists of two stages, precomputation and the runtime. They first precomputation included the vibration modes, consisting of mode shapes and frequencies. The authors used BEM solvers to estimate the boundary solution data for reconstruction of the acoustic transfer function. At runtime, they precomputed equivalent multipole sources to approximate each acoustics transfer equation for efficient synthesis. This method achieves adequate sounding effects for vibrating objects, but it still suffers from large amounts of precomputation. To this end, Wang and James \cite{Wang19} improved PAT and proposed KleinPAT, an optimal mode conflation method for time-domain precomputation of acoustic transfer. FFAT (far-field acoustic transfer) cube maps \cite{Jef09} were employed for efficient approximation of the transfer field. It was reported that KleinPAT achieved precomputation rates 2000x faster than the BEM-based approach. Li et al. \cite{Li15} proposed an interactive acoustic transfer approximation method specified for modal sound. Wang et al. \cite{Wang18} proposed a general, offline sound propagation method for a variety of physics-based simulation models. The key ingredient of this method is a sharp-interface finite-difference time-domain (FDTD) wave solver. A time-parallel sound synthesis method was also designed on commodity cloud computing resources. Note that this method can be used for computing general near field radiation, not just rigid bodies.

\textbf{Proxy-based methods.} Although it is possible to compromise between the speed and quality of modal sound synthesis to improve the speed of sound synthesis by using the special sparsity of modal sound in the frequency domain or the auditory perception ability of human ear, these methods are still limited to the synthesis of hundreds of objects in collision contact at the same time \cite{Rag06}. In most complex scenes, there may be thousands or even tens of thousands of objects colliding or breaking at the same time. Chadwick et al. \cite{Cha12} proposed a model to synthesize rigid-body acceleration noise due to objects undergoing short acceleration pulses. The essence of this method is to speed up the sound generation by calculating the \textsl{sound agent} in advance, but the method still needs to calculate the modal sound of different shapes of objects in advance.

\begin{figure}[!t]
\centering
\includegraphics[width=3.5in]{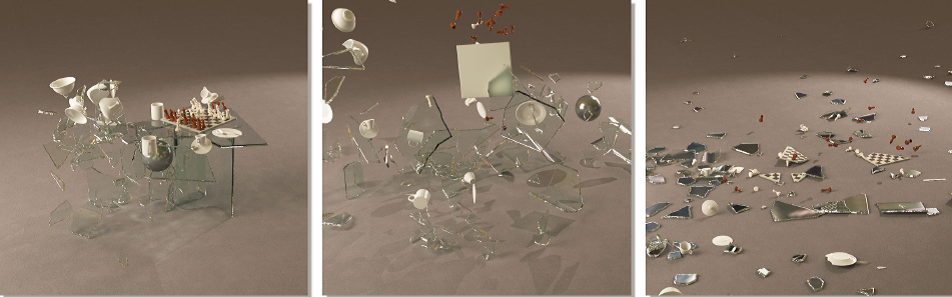}
\includegraphics[width=3.5in]{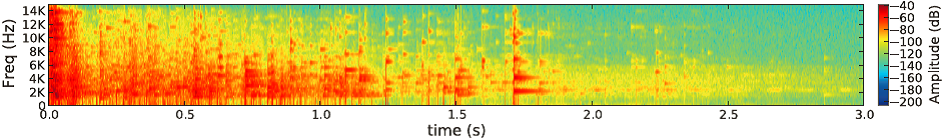}
\caption{The violent fracture and the synchronized impact sounds of a glass table setting smashed into over 300 pieces \cite{Zheng10}.}
\label{fig4}
\end{figure}

In order to reduce the calculation of a sound generation model in complex fracture scenes, Zheng and James \cite{Zheng10} proposed a method that uses a precomputed ellipsoid sound source proxy in their rigid-body fracture sound synthesis model. This method approximates different shapes of fragments to different sizes of ellipsoids, and precomputed a group of rigid-body sound banks, which greatly improves the sound synthesis of a fracture scene at animation speed. Figure \ref{fig4} shows a violent fracture and the synchronized impact sounds of a glass table getting smashed into over 300 pieces.  Chadwick et al. \cite{Cha12_2} applied the ellipsoid-based sound source proxy method to the sound synthesis of animations where many objects generate the acceleration noise at the same time, effectively improving the generation speed of acceleration noise. However, using sound proxy reduces the quality of synthesized sound, and it still takes tens of minutes to calculate the breaking sounds of a simple table. Although the calculation speed is much faster than methods without proxy, it still needs to be further improved to increase its practical performance.

\textbf{Event-driven granularity synthesis method.} Most of the methods of synthesizing breaking sounds use recordings and combine breaking events to synthesize the sound for breaking animation. In order to improve the utilization of recording, Picard et al. \cite{Pic09} recorded, analyzed, and recombined the sounds of object collision and breaking by using particle size synthesis technology before generating reasonable sound fragments for a rigid-body simulator. By using the output of the physical engine to select recording particles, the method automatically synthesizes the sound corresponding to different events (such as collision, fragmentation, and rolling). In early psychophysical experiments, Warren and Verbrugge \cite{War84} studied the bouncing and breaking sounds of glass bottles and the ability of listeners to correctly recognize the two sounds. Through artificial recognition and editing of impact/fracture sound events, credible sound fragments were synthesized, and it was determined that the listener was unable to  distinguish which fracture event was the source of fracture sound, thus simplifying the steps of simulating synchronous fracture events. In this paper, based on the method of particle size synthesis driven by a pulse event, the method of fracture event synthesis used the visual masking effect of explosion animation on a fracture event, and employed the method based on a single pulse-type response to synthesize a reasonable fracture sound.

\begin{figure*}[!t]
\centering
\includegraphics[width=7.0in]{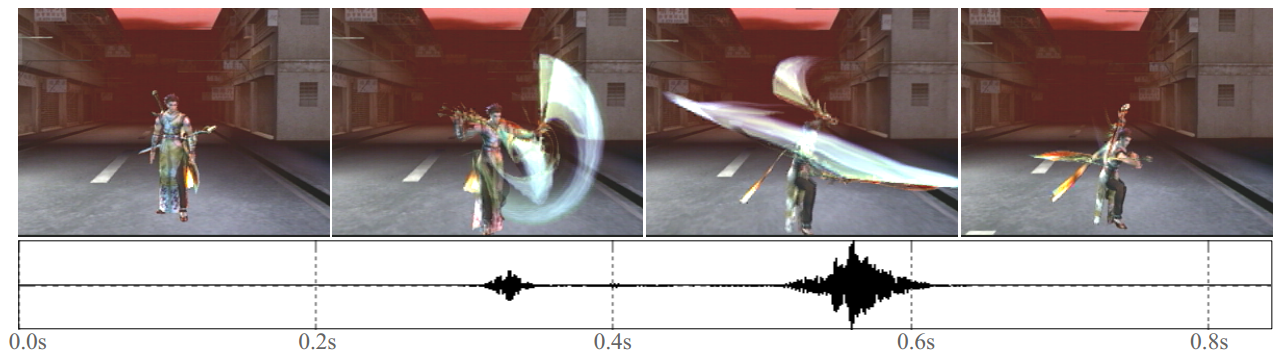}
\caption{Visual animation and waveform of the corresponding sound of a sword dance \cite{Dob03}.}
\label{fig12}
\end{figure*}

\subsubsection{Aerodynamic sound}
Aerodynamic sounds are generated by the flow of air around an object, i.e. oscillations in the pressure field given a particular geometry. Dobashi et al. \cite{Dob03} simulated two types of aerodynamic sounds, aeolian tones and cavity tones. The former tones are usually produced by stick-like objects presenting in a flow such as a sword waving in the air, and the latter are usually generated by the flow around hollows (or cavities) such as wind blowing through a window. Dobashi et al. \cite{Dob03} assumed that the main source of aerodynamic sound is an air fluid vortex. First, this method used computational fluid dynamics (CFD) to generate the sound texture of air sounds. Then, these sound textures were employed to synthesize aerodynamic sounds in real time according to the motion of the object or the speed of the wind. Because the method is simple and efficient, it represents a great contribution to the research of fluid sound at that time. It was successfully used in the synthesis of aerodynamic sounds such as those in a sword dance scene, as shown in Figure \ref{fig12}. 

Xu and Liu \cite{Xu20} proposed a sound synthesis method for rotating blades such as fans, helicopters, and wind turbines. They used the Ffowcs Williams-Hawkings (FW-H) equation to model the sound caused by moving objects, which consists of the thickness noise and the load noise. Here, the thickness noise is caused by continuous insertion or removal of a rotating blade of a certain thickness in the air, while the load noise is caused by the object moving in the air, meaning that it will be subjected to pressure from the surrounding air. They extracted Mel-scale Frequency Cepstral Coefficients (MFCCs) from the sound produced by the above physically-based synthesis result and then convolved the MFCCs with the white noise to enrich the rotating blade sound. Li et al. \cite{Li16} proposed a method that can automatically design acoustic filters with complex geometries.

Wind instruments also produce sound through air vibration. These instruments play an important role in music and show great diversity in structure and sound. However, most wind instruments have cylinder shapes, and the overall acoustic model is relatively simple. The acoustic simulation of other shapes is usually complex. Gordon \cite{Gor00} used a sine wave to generate sound, which exploits harmonics of different frequencies and amplitudes to synthesize the sound of various musical instruments, and the calculation is fast. Based on this foundation, Allen et al. \cite{All15}  proposed a real-time simulation method to generate full-bandwidth sound for 2D wind instruments. This method formulated wind instrument sound synthesis as solving a wave field on a 2D instrument geometry. The high computational efficiency is facilitated by a novel 2D wave solver based on the Finite-Difference Time-Domain (FDTD), which simulates audio at $128,000$Hz. Umetani et al. \cite{Ume16} introduced a simulation method of 3D free-form musical wind instruments. Compared with the general acoustic simulation, the biggest difference is that it is not meant to solve the whole spectrum, but only focuses on the dominant sound, which is the lowest passive resonance frequency of the inner cavity of the musical instruments. This method can solve for the unusual shape of wind instruments and generate sound in real time for user interaction, i.e. to solve the generalized eigenvalue problem from a group of sparse acquisition in the frequency domain.

\begin{figure}[!t]
\centering
\includegraphics[width=3.5in]{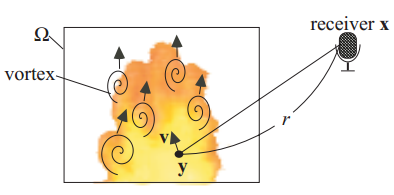}
\caption{Vortex noise in a fire \cite{Dob04}. Note that $\Omega$ denotes the turbulent field, $y$ is a point in $\Omega$, $r$ is the distance between the receiver and the point $y$, $v$ is the velocity of $y$.}
\label{fig5}
\end{figure}

\subsubsection{Flame sound}  Dobashi et al.~\cite{Dob04} generated flame noises from a fire animation model. This method focuses on the complex motion of the vortex in a fire, and uses the computational fluid dynamics method to calculate the vortex to simulate the vortex noise. Figure \ref{fig5} is the schematic diagram of the vortex noise. The vortex sound is computed in the preprocessing stage by sound texture, and the real-time flame sound is synthesized by sound texture. This paper shows some examples of generating sound through turbulent flames. Although these examples demonstrate the diversity of aerodynamic sound models, subsequent studies show that the main source of combustion sound is not vortex-based noise. Experiments \cite{Sch09}\cite{Ihm09}\cite{Chr92} showed that, for most flame types, another sound source (i.e. direct combustion sound) is the main contributor of combustion noise, and aerodynamic noise only plays a relatively small role.

Nguyen et al.~\cite{Ngu02} used two sets of independent Navier-Stokes equations to simulate the flame motion and used the level set method to track the interface at the front of the flame. Meanwhile, the image mirror value method was used to meet the different jumping conditions of pressure and speed at the interface. The numerical method based on decomposition thermoacoustic chemical reaction flow in physics can also generate combustion noise. These purely physics-based methods can get accurate flame data that can be used for flame sound synthesis. However, such simulation methods (such as large eddy current simulation) are several orders of magnitude more expensive than visual flame simulation models in CG and the time consumption is huge, which severely limits the practicality of sound synthesis. Therefore, purely physics-based methods are not suitable for reproducing flame sounds of different combustion media.
 
\begin{figure}[!t]
\centering
\includegraphics[width=3.5in]{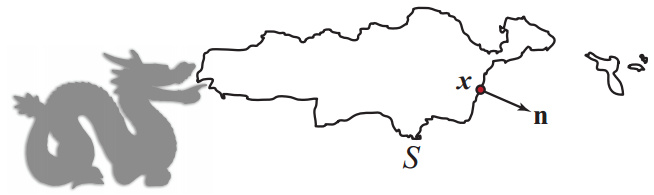}
\caption{Illustration of the surface of the flame front end \cite{Cha11}. Note that $S$ denotes the flame front, turbulent field, $n$ is the normal of point $x$.}
\label{fig6}
\end{figure}

Chadwick and James ~\cite{Cha11} combined noise bandwidth expansion and sound texture synthesis to synthesize the sound of flames. They divided the sound into low-frequency and high-frequency parts. For the low-frequency part, the physically-based flame simulation ~\cite{Ngu02}\cite{Hon07}\cite{Hor09} was used as input, the fuel value data of the flame front end was obtained, and then the data was fed into the sound wave equation to evaluate the low-frequency sound signal, as shown in Figure \ref{fig6}.

The bandwidth expansion technique was exploited to calculate the high-frequency part of the sound. Marelli et al.~\cite{Mar10_2} used the spectral noise synthesis method to synthesize the sound fragments of flames, and they found that the lost time structure can be modeled with higher-level statistical information. In comparison, Chadwick and James~\cite{Cha11} used the spectral noise model and obtained the time structure from the flame sound based on low-frequency information. Since there is a certain power-law relationship when the signal frequency reaches a certain peak value~\cite{Abu78}, Chadwick and James~\cite{Cha11} randomly generated the high-frequency sound signal of the flame according to the power law. With texture synthesis, the specific flame sound samples were segmented according to the fixed window size, and the low-frequency sound signals were filtered to obtain the low-frequency signal waveform pyramids with different level of  accuracy. Then the nearest neighbor search was carried out according to the segmented window size to find the corresponding sound texture, and the final flame sound signals were derived by fusion. 

Liu and Yu~\cite{Liu15} proposed flame sound synthesis method by further extending the bandwidth expansion method. This method adapts the Fourier transform and inverse transform to process the low-frequency signal, uses wavelet noise to synthesize the high-frequency information, then mixes the two parts to get the final flame sound. Liu and Yu's method~\cite{Liu15} reduced the running time, as compared to~\cite{Cha11}.

\begin{figure*}[!t]
\centering
\includegraphics[width=6.5in]{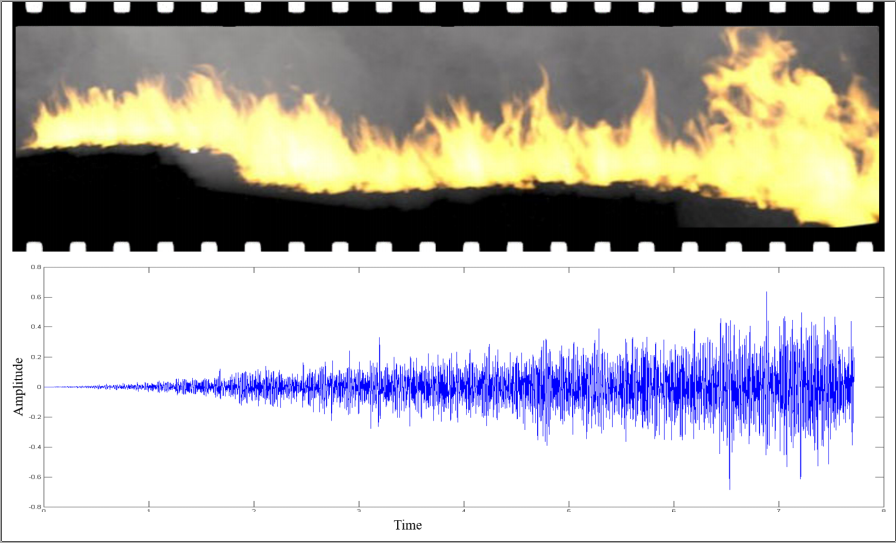}
\caption{Visual simulation of a fire spreading and the corresponding sound waveform \cite{Yin18}.}
\label{fig7}
\end{figure*}

The above methods can produce plausible flame sounds that are synchronized with the flame animation, however, these methods do not distinguish between different combustion media (e.g., the sound of paper burning is quiet, while with carbon and wood, a "bang" and a "crackle" can be heard, respectively). Yin and Liu~\cite{Yin18} proposed an interactive fire sound synthesis method by considering the influence from different solid combustibles. This paper chose the non-premixed flame animation model as input. The direct combustion noise, the turbulence vortex noise, and the interaction noise were regarded as the main sound sources. For the direct combustion noise, they used the marching-cube-like method~\cite{Liu15} to model the front end of the flame; for the turbulence vortex noise, they used the sub-grid method to model the turbulence in each voxel to get the turbulence vortex noise; for the interaction sound, they proposed a modified empirical mode decomposition method (EMD) to extract and fuse the interaction sound details to the sound of flame with the interaction sound. Meanwhile, the high-frequency flame segment details were produced based on high-frequency hybrid segment matching to add high-frequency flame details to the flame sound, making the final flame sound richer and more real. Figure \ref{fig7} shows the sound results for a spreading fire.

\subsubsection{Explosion sound}
Explosion animation is one of the most common scenes in entertainment such as games and special effects. The synthesis of explosion sound synchronized with an explosion animation was not adequately solved until recently. The explosion sound consists of explosive sound and combustion noise. The explosive sound is generated by a large amount of released energy, which is a brief and intense transient sound accompanied by the phenomenon of a fireball in an explosion. After the big bang of the fireball, there will be a duration of combustion, which produces combustion noise. In addition, when the explosion interacts with objects of different materials, the objects also break and thereby cause explosive sounds. It is necessary to also consider the sound generated by cracking and fracture in explosion \cite{Liu20_3}.

\begin{figure}[!t]
\centering
\includegraphics[width=3.5in]{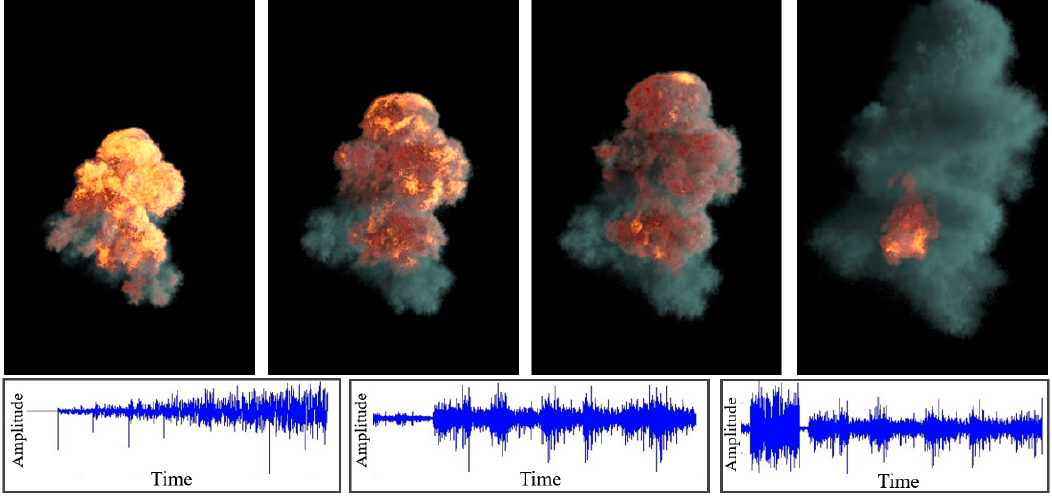}
\caption{The results of the synthesized smoke explosion sound. The top row: the visual smoke explosion model including fireball, burning, and diffused smoke; the bottom row (from left to right): the sound waveforms of the synthesized low-frequency explosion sound, the combustion noise, and the resulting explosion sound \cite{Liu20}.}
\label{fig13}
\end{figure}

An automatic generation method was proposed \cite{Liu20} for realistic sound simulation of an explosion animation. This paper mainly studied the sound corresponding to the two main phenomena in the process of explosion, that is, the explosive sound corresponding to the fireball phenomenon and the combustion noise corresponding to the combustion phenomenon. Based on the sound samples and the relevant parameters derived from the physical animation, the two kinds of sound were synthesized. To synthesize explosive sound, the occurrence and duration of explosive sound are determined according to the dynamic change of fuel consumption during the production of explosion animation, so that the explosive sound matching the fireball animation can be extracted from the recording samples according to the high-frequency content. To synthesize combustion noise, a method based on the similarity of timbre was proposed. According to the timbre characteristics of low-frequency explosion sound synthesized by physical methods, they selected sound particles similar to their timbre from the constructed sound corpus and synthesized the complete combustion noise. Finally, according to the occurrence and duration of the explosive sound, the two sounds were mixed, and a reasonable explosion sound was produced. Figure \ref{fig13} demonstrates the results of the synthesized smoke explosion sound.

\subsubsection{Liquid sound} 
Liquid sound is primarily caused by bubble formation and resonance. Previously, researchers mainly studied the physical model of water sound generation, that is, the acoustics of bubbles. For example, Strasberg~\cite{Str53} described the influence of bubble shape and its surroundings on bubble sound and generalized this theory to non-spherical bubbles. In 1994, Leighton et al.~\cite{Lei94} proposed that the formation and resonance of bubbles are the most important sources of water sound. This research has become the theoretical basis of sound generation in many water scenes. For all kinds of water scenes, it is found that ocean noise~\cite{Pro88}, rain sound~\cite{Lon90}, and even the sound generated during whale hunting~\cite{Lei07} can be approximated by the acoustic bubble model. The acoustic wave solution formula of a spherical, isolated bubble can be expressed as:
\begin{equation}
p(t)=A \sin(2 \pi ft)e^{-\beta t},
\end{equation}
where $f$ is the resonance frequency of a single bubble, $\beta$ is the damping term, $A$ is the bubble amplitude, and $t$ is time. Note that, except for the variable $t$, the values of other parameters are related to the bubble radius.

Van Den Doel~\cite{Doe05} proposed an isolated single bubble sound model and verified its feasibility through a small-scale user survey. Van Den Doel also designed a stochastic model to synthesize interactive and complex liquid sound in real time, e.g., pouring water, a river, and rain. Users can manually adjust the bubble details and the number of parameters to modify the fluid sound in complex scenes.

Deane~\cite{Dea13} performed a study of the sound produced by bubbles (2.5 mm radius) bursting on the surface of water and found that the sound pulse was related to the time interval between the bubble reaching the water surface and breaking. The bubble will burst and emit a chirped pulse sound in 10 milliseconds, just like the wave of a Helmholtz resonator. The bubbles that lasted for more than $100$ milliseconds exhibited more complex acoustic behaviors than just popping. Deane analyzed the behavior of the resonators and estimated the relationship between the film thickness and bubble resonance. Although this method could generate good shock waves or bursting sounds, that may not be sufficient for underwater sounds. The sound of water flow movement is generated by bubble vibrations, not by burst sound. As a result, this method is only suitable for modeling water flow sound with splashing liquids.
 
Guo and Williams~\cite{Guo91} deduced a mathematical expression to represent the impact sound with an initial period of less than $1$ microsecond. Howe and Hagen~\cite{How11} proposed using impact sound to synthesize the sound of droplets falling into water. Howe and Hagen's work extended the activity cycle to the whole effective life cycle (about 100 microseconds) on the basis of method proposed by Guo and Williams~\cite{Guo91}. Although this method can produce the sound of droplets falling into water, it does not conform to the actual underwater sound model. Sometimes, the sounds of objects falling into the water come from bubble vibration after collision, rather than the sound of the contact body. Considering the collision sound alone is not applicable in some water models without collision, such as streams and rivers.

\begin{figure}[!t]
\centering
\includegraphics[width=3.5in]{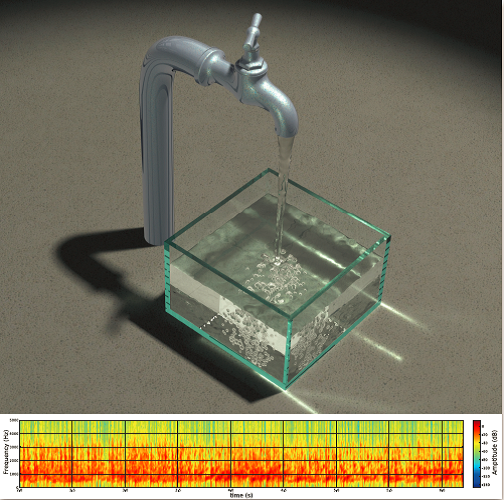}
\caption{Synthesizing the sound of pouring water via the linear superposition of acoustic radiation from 7900 vibrating acoustic bubbles \cite{Zheng09}.}
\label{fig8}
\end{figure}

In recent years, most of the liquid acoustic research has been based on the generation of harmonic bubbles. The earliest work is a type of bubble-based fluid sound method~\cite{Imu07}. This work pointed out that the generation of underwater sound is closely related to the diameter and movement of bubbles in water. Based on the SPH fluid simulation~\cite{Kel06}, they used the data sample method to enhance the sound of SPH fluid simulation based on the independent bubble recording. This method can not capture the structural information of three-dimensional sound radiation in time-varying space. Zheng and James~\cite{Zheng09} added particles to the existing fluid simulation methods and simulated the generation, vibration, advection, and radiation of bubbles through the motion evolution of these particles, thus avoiding audio-rate time-stepping of fluids and achieving real-time sound synthesis. They used a set of acoustic propagation functions from bubbles to human ears to measure the vibration of bubbles. In this paper, the discrete Green function based on the Helmholtz equation was used to calculate the oscillation frequency, and it was applied to the specific sound propagation. In order to solve the Helmholtz equation derived from a large number of bubbles in real time, a fast dual-domain multipole boundary-integral solver was designed. This method can synthesize liquid sound synchronized with animation; however, it is not suitable for multiphase flow since the simulation time is in hours. At the same time, this method only synthesizes the sound of bubbles and ignores the sound produced by water interaction. 

\begin{figure}[!t]
\centering
\includegraphics[width=3.5in]{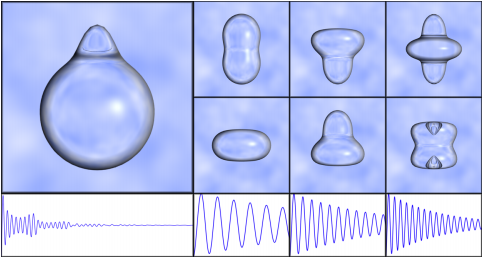}
\caption{Illustration of a simple bubble decomposing into spherical harmonics. The upper left shows the original bubble. The two rows on the upper right show the two octaves of the harmonic deviations from the sphere. Along the bottom is the sound generated by the bubble and the components for each harmonic \cite{Mos10}.}
\label{fig9}
\end{figure}

Moss et al. \cite{Mos10} proposed a modeling method based on physical harmonic bubbles. It is believed that the generation of fluid sound is mainly due to the vibration of bubbles resonating in the medium. They coupled a real-time shallow water model and a bubble equation based on physics to develop an automatic underwater acoustic synthesis algorithm. By considering both the spherical and non-spherical bubble shapes (see Figure  \ref{fig9}), they proposed a way to adaptively select how many modes to use for a bubble and provided a mechanism to evaluate the sound produced by either single- or multi-mode bubbles based on the Leighton’s bubble theory  \cite{Lei94}. For the spherical bubble, the attenuation sine function was used to solve the sound. For the non-spherical function, based on the  Higgins theory \cite{Lon89}\cite{Lon89_2}, the non-spherical attenuation curve was derived on the basis of the attenuation sine function. This method considers the influence of different forms of bubbles in the physical level, which makes the resulting sound more appealing.

Langlois et al. \cite{Lan16} synthesized the sound of bubbles based on physics from two-phase incompressible fluids. By tracking the geometry of the contact surface between the fluid and air, it can directly identify the geometry and topological results of the bubbles, which are changed due to entrainment, merging, and spitting. They proposed a novel capacitance-based method to estimate the change of bubble frequency caused by bubble size and bubble shape, as well as the bubble's proximity to rigid and free surfaces. They leveraged a set of bubble forcing models \cite{Dean08}\cite{Czer11}\cite{Czer11a} to compute the entrainment, splashing, and fusion of bubbles. In order to overcome the limitation of frequency bandwidth caused by the resolution of the fluid grid, they proposed simulating micro-bubbles in the sound domain, which adopted the power-law model of the overall distribution of bubbles. This method can synthesize plausible underwater sound by adding bubble position and sound propagation. Wang et al. \cite{Wang18} further proposed a novel wave-based sound synthesis method for physics-based simulation models and computer animations. This method can achieve high-quality off-line sound synthesis effects by resolving the sound wave equation with near-field scattering and diffraction.

The above methods can handle the sound generation of small-scale scenes very well, but most of them are not suitable for large-scale water scenes such as ocean waves. Means et al. \cite{Mea01} proposed that the low-frequency sound of breaking waves in the open sea can be generated by modeling and calculating the resonance of bubble clouds in the waves and gave the corresponding hemispherical model. They \cite{Mea04} further extended the theory of bubble cloud resonance to the generation of surf sounds by adjusting the hemispherical model to the long leaf semi-cloud model. However, the physical mechanism of this kind of model is complex and cannot be applied to the sound generation of an ocean wave scene directly, either in terms of algorithm efficiency or application difficulty.

Wang and Liu \cite{Wk18} proposed a hybrid method to generate the sound of wave animation based on available wave sound samples and bubble extraction. Based on the theory of bubbles as the main sound source, they generated bubble particles to ensure the synchronization between visual and sound effects. In the simulation stage, according to the different depths of the fluid particles, some of them were divided into bubble particles. Then, the position and velocity information of these bubble particles was extracted from the animation. For each animation frame, they used a clustering algorithm to divide the bubble particles mapped to the two-dimensional plane into different wave regions, thus forming the wave division in each frame. For each wave area, they extracted the total velocity of the bubble particles as the attributes of the wave. Finally, a sound generation method based on loudness was used to generate the sound of waves using sound samples. In the whole process, a nonlinear mapping function was employed to associate the particle attributes with the sound clip.

\subsubsection{Rigid-fluid interaction sound}

The above methods can recreate fine liquid sound, but most of them neglect the liquid-solid interaction that produces a characteristic sound apart from the pure sound of bubbly water flows. Wilson et al. \cite{Wil17} presented an effective approach to simulate the sound of non-empty objects containing fluids. They modified the conventional sound synthesis equation by introducing the fluid force on an object at the fluid-structure boundary. 

\begin{figure}[!t]
\centering
\includegraphics[width=3.5in]{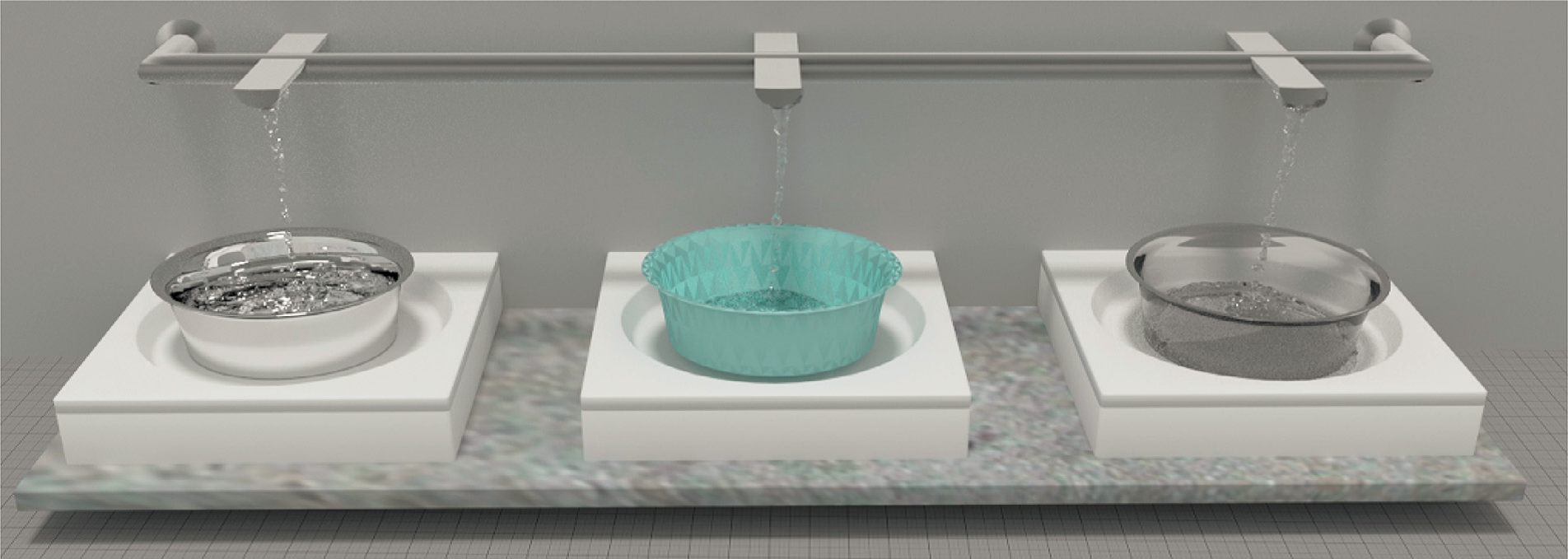}
\caption{An example of interaction sound between pouring faucet and basins of different materials \cite{Cheng19}.}
\label{fig10}
\end{figure}

Cheng and Liu \cite{Cheng19} proposed a liquid-solid interaction sound (LSIS) synthesis method, that includes two components, namely bubble sounds and impact sounds corresponding to liquid-liquid collision and liquid-solid collision, respectively. A sound enrichment method called Feature Transfer Synthesis (FTS) was designed to enrich the synthesized sound. Figure \ref{fig10} shows an example of interaction sounds generated from a pouring faucet and basins of different materials.


In Virtual Reality (VR) application, efficiency plays an important role. It is challenging to achieve interactive rigid-fluid interaction sound synthesis. Cheng and Liu \cite{Cheng19_1} proposed a novel sound synthesis framework specific to VR systems. They took advantage of information from the haptic channel and designed a haptic force-guided granular sound synthesis method. A multi-force (MF) granulation algorithm was developed to balance the algorithm efficiency and synchronization. It was reported that this was the first attempt to introduce the tactile information for rigid-fluid interaction sound synthesis.

\subsubsection{Rain sound}
There are more than tens of thousands of raindrops in a rainfall scene, all of which are sound sources. It is very challenging to synthesize such sounds using traditional physics-based sound synthesis methods. Tan \cite{Tan90} proposed rain sound characteristics in a heavy rainfall and showed a good correlation between raindrop velocity and sound level. Medwin et al. \cite{Med92} pointed out that when raindrops fall into larger bodies of water, they will produce underwater sound in a wide range of acoustic frequencies. They tested more than 1000 single raindrop samples, covering most of the raindrop size range, and proved that raindrops can be classified by four obviously different radius values and that raindrop sound can be divided into collision sound and microbubble sound. Van Den Doel et al. \cite{Doe04} presented a method for synthesizing sound produced by scenes with many sounding objects. They realized fast audio rendering by exploiting modal synthesis and efficiently eliminating the inaudible modes according to the masking characteristics of the human auditory system. They demonstrated a rainfall scene on a roof. 

\begin{figure}[!t]
\centering
\includegraphics[width=3.0in]{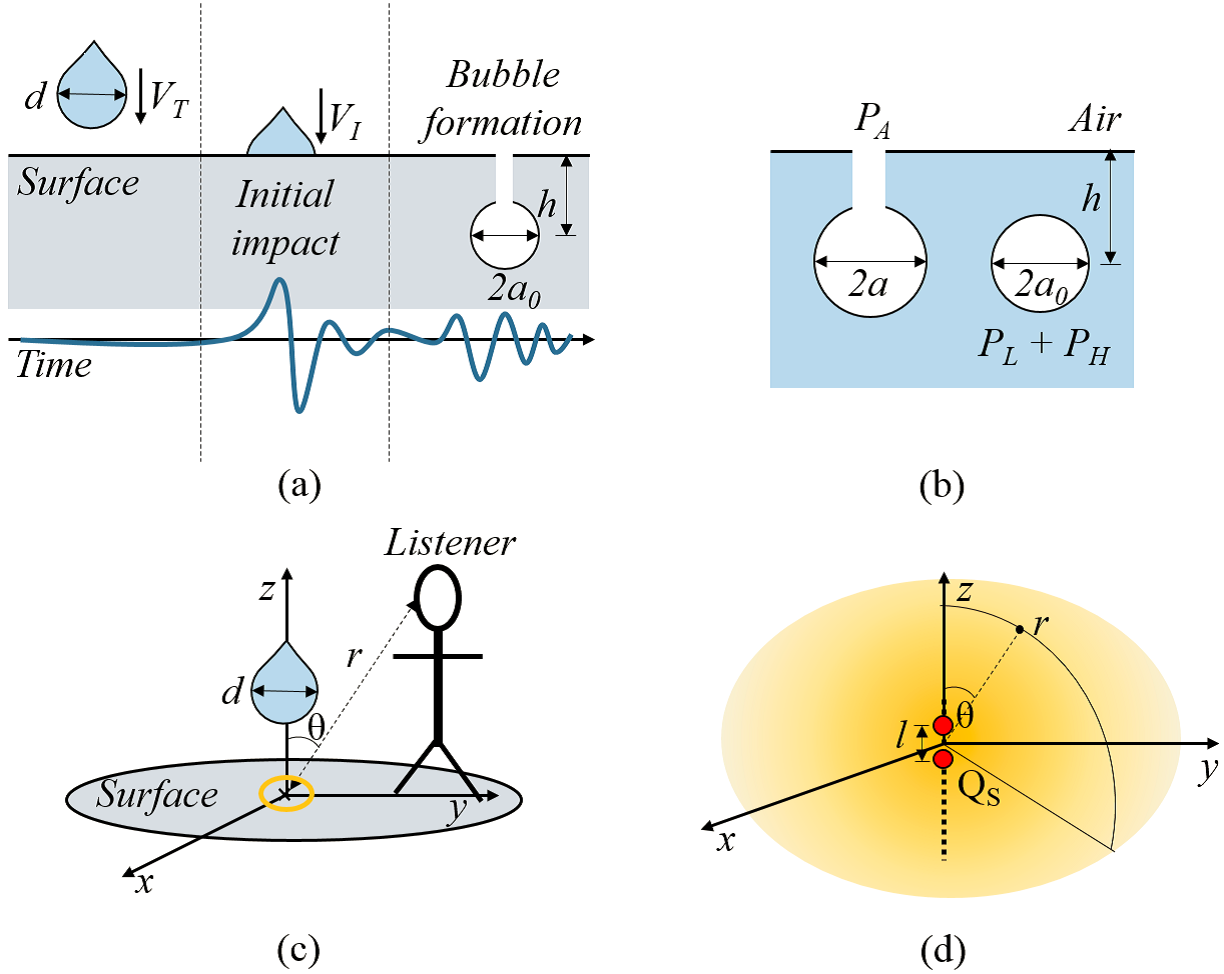}
\caption{Models for a raindrop. (a) the process of a raindrop impacting the surface and the corresponding acoustic wave patterns, (b) the variation of the entrained bubbles, (c) a listener model, and (d) a close-up view of the orange ellipse in (c) \cite{Liu19}. Note that $d$ is the raindrop diameter, $r$ is the distance from a listener to the sound source, $\theta$ is the polar angle, $V_T$ is the terminal velocity of the raindrop, and $Q_S(t)$ illustrate the dipole sound field geometry. For more detais, please refer to \cite{Liu19}.}
\label{fig11}
\end{figure}

Recently, Liu et al. \cite{Liu19} proposed a physically-based statistical simulation method for rain sound. Figure \ref{fig11} shows models for a raindrop. The initial impact and the subsequent pulsation of entrained bubbles were considered to model the raindrop sound. The material sound textures (MSTs) were constructed to distinguish rain sounds on different surface materials. A basic rain sound (BR-sound) bank was built offline according to the raindrop statistical model. This method drastically decreases the computational cost compared to traditional physics-based sound synthesis methods.

\subsubsection{Thin-shell sound}
\begin{figure}[!t]
\centering
\includegraphics[width=3.5in]{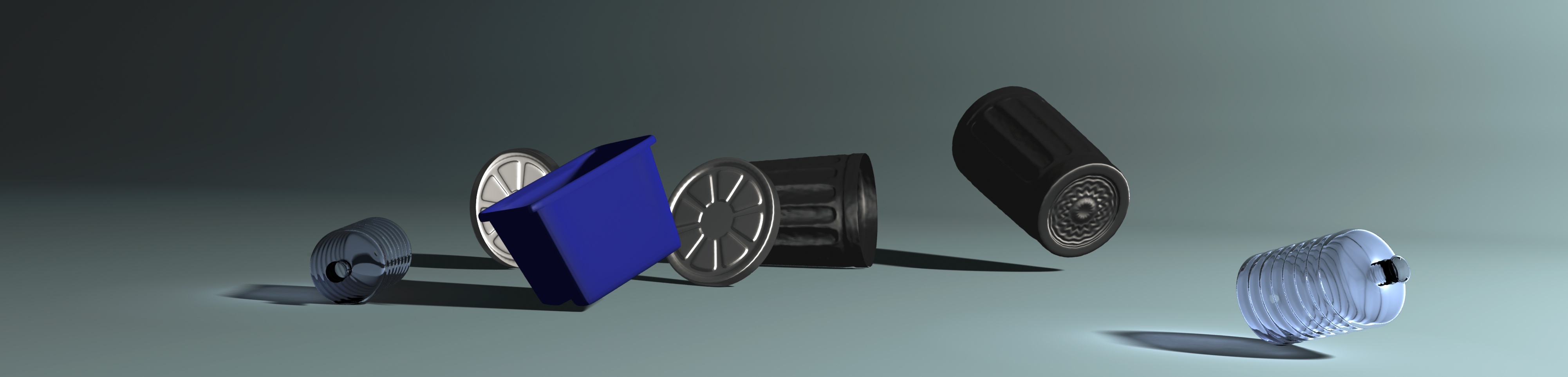}
\caption{ Physically-based sound renderings of thin shells \cite{Jef09}.}
\label{fig14}
\end{figure}

Thin-shell objects with thicknesses much smaller than their other dimensions such as trash cans, oil drums, and tin roofs are common, and they produce distinct impact sounds. Chadwick et al. \cite{Jef09} proposed a practical nonlinear sound model (see Figure \ref{fig14}), harmonic shells, for near-rigid thin-shells. Linear modal analysis was adapted to generate a small-deformation displacement basis. They estimated high-resolution sound field approximations with far-field acoustic transfer (FFAT) maps in pre-computation, for efficient estimation of low- to high-frequency acoustic transfer functions. This method achieves satisfactory sounding results for nearly rigid thin shells that vibrate under small deflections.

\begin{landscape}

\begin{table}
\centering
\caption{Sound synthesis methods.}
\label{table1}

\begin{tabular}{ccc}

\hline
\specialrule{0em}{5pt}{5pt}
\multicolumn{2}{c} {Methods}                           & References           \\  
\specialrule{0em}{5pt}{5pt}
\hline 
\specialrule{0em}{5pt}{5pt}
\multicolumn{2}{c} {Harmonic synthesis methods}              & Gordon 2000 \cite{Gor00}              \\ 
\specialrule{0em}{5pt}{5pt}

\multicolumn{2}{c} {Texture synthesis methods}               & \tabincell{c} {Dubnov et al. 2002 \cite{Dub02}, Cardle et al. 2003 \cite{Car03}, Ren et al. 2013 \cite{Ren13}, \\ Sterling and Lin 2016 \cite{Ste16}, Zhang et al. 2019 \cite{Zhang19} }             \\ 
\specialrule{0em}{5pt}{5pt}

\multicolumn{2}{c} {Spectral analysis methods}               & Serra and Smith 1990 \cite{Ser90}, Marelli et al. 2010 \cite{Mar10}, McDermott et al. 2009 \cite{Mcd09}              \\
\specialrule{0em}{5pt}{5pt}
\multirow{11}{*}{Physics-based synthesis methods} & Rigid-body sound synthesis methods         & \tabincell{c} {Van Den Doel et al. 2001\cite{Doe01}, O$^\prime$Brien et al. 2002 \cite{Obr02}, Raghuvanshi and Lin 2006 \cite{Rag06}, \\ James et al. 2006 \cite{Jam06}, Raghuvanshi et al. 2008 \cite{Rag08}, Bonneel et al. 2008 \cite{Bon08}, \\ Zheng and James 2010 \cite{Zheng10}, Ren et al. 2010 \cite{Ren10}, Zheng and James 2011 \cite{Zheng11}, \\ Chadwick et al. 2012 \cite{Cha12}, Chadwick et al. 2012 \cite{Cha12_2}, Langlois and James 2014 \cite{Lan2_14}, \\ Langlois et al. 2014 \cite{Lan14}, Ren and Lin 2015 \cite{Ren15}, \\ Li et al. 2015 \cite{Li15},  Schweickart et al. 2017 \cite{Sch17}, Wang et al. 2018 \cite{Wang18}, Sterling et al. 2019 \cite{Ste19}  }\\
\specialrule{0em}{5pt}{5pt}

  & Flame sound synthesis methods         & \tabincell{c} { Dobashi et al. 2004 \cite{Dob04}, Chadwick and James 2011 \cite{Cha11}, \\ Liu and Yu 2015 \cite{Liu15}, Yin and Liu 2018 \cite{Yin18}} \\
\specialrule{0em}{3.5pt}{3.5pt}

  & Liquid sound synthesis methods         & \tabincell{c} { Van Den Doel 2005 \cite{Doe05}, Imura et al. 2007 \cite{Imu07}, Zheng and James 2009 \cite{Zheng09}, \\ Moss et al. 2010 \cite{Mos10},  Langlois et al. 2016 \cite{Lan16}, Wang and Liu 2018 \cite{Wk18}} \\
 \specialrule{0em}{3.5pt}{3.5pt}
  
  & Rigid-fluid interaction sound synthesis methods        &  Wilson et al. 2017 \cite{Wil17}, Cheng and Liu 2019 \cite{Cheng19}, Cheng and Liu 2019 \cite{Cheng19_1} \\
 \specialrule{0em}{3.5pt}{3.5pt}
   
  & Rain sound synthesis methods         &  Liu et al. 2019 \cite{Liu19} \\
 \specialrule{0em}{3.5pt}{3.5pt}
    
  & Aerodynamic sound synthesis methods         &   Dobashi et al. 2003 \cite{Dob03}, Allen and Raghuvanshi 2015 \cite{All15},\\ Umetani et al. 2016 \cite{Ume16}, Xu and Liu 2020 \cite{Xu20} \\
  \specialrule{0em}{3.5pt}{3.5pt}
  
  &  Explosion sound synthesis methods         &   Liu and Gao 2020 \cite{Liu20} \\
\specialrule{0em}{3.5pt}{3.5pt}
   
  &  Cloth sound synthesis methods             &   An et al. 2012 \cite{An12} \\
  \specialrule{0em}{3.5pt}{3.5pt}
    
  & Thin-shell sound synthesis methods         &   Jeffrey et al. 2009 \cite{Jef09}, Cirio et al. 2016 \cite{Cir16}, Cirio et al. 2018 \cite{Cir18} \\
 \specialrule{0em}{3.5pt}{3.5pt}
  
  & Paper sound synthesis methods               &   Schreck et al. 2016 \cite{Sch16} \\
 \specialrule{0em}{3.5pt}{3.5pt}
  
  &  Footstep sound synthesis methods         &   	Nordahl et al. 2011 \cite{Nor11}, Turchet 2016 \cite{Tur16} \\
\specialrule{0em}{5pt}{5pt}
  
  \multicolumn{2}{c} {Audio generations methods for videos}               & Li et al. 2018 \cite{Li18}, Owens et al. 2016 \cite{Owe16}, Zhou et al. 2018 \cite{Zhou18}, \cite{Meh17}, \cite{Cheng19_3}              \\
\specialrule{0em}{5pt}{5pt}
  
  \multicolumn{2}{c} {Deep audio synthesis methods}               & Jin et al. 2020 \cite{Jin20}              \\
\specialrule{0em}{5pt}{5pt}
  
\hline
\end{tabular}
\end{table}

\end{landscape}

\begin{table*}[htb]
\centering
\caption{Sound propagation methods}
\label{table2}
\setlength{\tabcolsep}{0.36in}
\begin{tabular}{cc}
\hline
\specialrule{0em}{3.5pt}{3.5pt}
Methods                          & References           \\  
\specialrule{0em}{3.5pt}{3.5pt}
\hline 
\specialrule{0em}{3.5pt}{3.5pt}
Wave-based methods                & \tabincell{c} { Botteldooren 1994 \cite{Bottel94}, Savioja et al. 1994 \cite{Savioja94}, Botteldooren 1995 \cite{Bot95}, \\ James et al. 2006 \cite{Jam06}, Raghuvanshi et al. 2009 \cite{Rag09}, Gumerov and Duraiswami 2009~\cite{Gum09}, \\ Raghuvanshi et al. 2010\cite{Rag10_2}, Mehra et al. 2012~\cite{Meh12}, Mehra et al. 2013~\cite{Meh13}, \\ Raghuvanshi and Snyder 2014 \cite{Rag14}, Mehra et al. 2014 \cite{Meh14}, Mehra et al. 2015~\cite{Meh15}, \\ Raghuvanshi and Snyder 2018 \cite{Rag18},  Morales et al. 2015~\cite{morales2015parallel}, Zhang et al. 2018 \cite{Zhang18}, \\ Chakravarty et al. 2019 \cite{Cha19}, Wang and James 2019 \cite{Wang19}, Liu and Liu 2020 \cite{Liu20_2}   }            \\ 
\specialrule{0em}{3.5pt}{3.5pt}
Geometric methods                & \tabincell{c} {Krokstad et al. 1968 \cite{krokstad1968}, Allen et al. 1979 \cite{allen1979}, Borish 1984 \cite{borish1984}, \\ Vorl{\"a}nder 1989 \cite{vorlander1989}, Funkhouser et al. 1998 \cite{Fun98}, Funkhouser et al. 1999 \cite{Fun99}, \\ Tsingos et al. 2001 \cite{Tsi01}, Siltanen et al. 2007 \cite{siltanen2007room}, Lauterbach et al. 2007 \cite{Lau07}, \\ Chandak et al. 2008 \cite{Cha09}, Tsingos et al. 2009 \cite {tsingos2009precomputing}, Laine et al. 2009 \cite{laine2009accelerated}, \\ Antani et al. 2011 \cite{antani2011direct}, Taylor et al. 2012 \cite{Mic12}, Antani et al. 2012 \cite{antani2012interactive}, \\ Schissler 2014 \cite{Schi14}, Mo et al. 2015 \cite{Mo15}, Schissler and Manocha 2016 \cite{Schi16}, \\ Cao et al. 2016 \cite{Cao16}, Schissler and Manocha 2016 \cite{Schi16_2}, Mo et al. 2017~\cite{Mo17} }           \\ 
\specialrule{0em}{3.5pt}{3.5pt}
Hybrid methods                   & \tabincell{c} { Yeh et al. 2013~\cite{Yeh13}, Rungta et al. 2016~\cite{Run16}, Rungta et al. 2018~\cite{Run18}, \\Schissler and Manocha 2018 \cite{Schi18}, Tang et al. 2020 \cite{Tang20_1} }\\ 
\specialrule{0em}{3.5pt}{3.5pt}
\hline
\end{tabular}
\end{table*}

Cirio et al. \cite{Cir18} proposed a multi-scale reduced simulation method for simulating \textsl{nonlinear} thin-shell sounds. They split nonlinear vibrations into two scales, namely a small low-frequency part and a high-frequency part. This method is very efficient and can reproduce shell deformation tens of times faster than traditional methods.

When crumpling a buckling thin-shell, gross elastic and plastic deformation are involved, creating distinct clicking sounds corresponding to buckling events. Cirio et al. \cite{Cir16} synthesized thin-shell sound through substructured modal analysis and stochastic enrichment. This method consists of four steps, i.e. buckling detection, modal vibration, sound enrichment, and sound propagation. High-quality buckling sound that is seamlessly synchronized with the animation can be produced in practical time. 

\subsubsection{Cloth sound}
An et al. \cite{An12} generated the sounds for cloth animation by dividing cloth sound into friction sounds and crumpling sounds. By artificially matching animation features with sound fragments, they established a mapping relationship between them. Then the selected sound fragments were spliced to generate the final sound results.

\subsubsection{Paper sound}
Schreck et al. \cite{Sch16} classified paper sounds into friction and crumpling. In contrast to prior methods, when these researchers set up a sound bank, they marked the sound clips according to the different states of the paper. After obtaining the animation information, according to the paper shape, the final result was generated by selecting the friction or kneading sound matching the state in the sound bank, and the spatial processing of the sound was also added in the process of generating the friction sound. 

\subsubsection{Footstep sound}
Walking sounds are particularly important for enhancing the immersiveness of computer games and virtual reality systems. The sounds of footsteps from different people walking on surfaces of different materials vary greatly. Cook \cite{Coo97} pioneered in this field by proposing a collection of physically informed stochastic models (PhiSM) \cite{Coo02} and simulating the various footsteps while considering different people and different materials. The above methods achieve plausible footstep sound results but may fail for some cases. However, these methods need tiresome parameter tweaking. Nordahl et al. \cite{Nor11} developed a footstep sound synthesis system, allowing a designer to synthesize sounds of different materials in real time. This method works independently from footwear and has great potential to be used in computer games where a user is able to navigate. Although this method works well for footstep sounds while walking on a flat surface, it pays little attention to the cases of uphill and downhill movements. Turchet \cite{Tur16} developed a footstep sound synthesis method based on physically based models
 combining additive synthesis and signals multiplication.

Table \ref{table1} summarizes the popular sound synthesis methods.

\section{Sound propagation}
\begin{figure}[!t]
\centering
\includegraphics[width=2.5in]{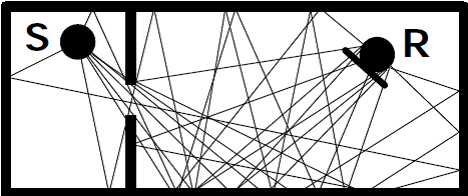}
\caption{Sound propagation paths from a source (S) to a receiver (R) \cite{Fun03}.}
\label{fig16}
\end{figure}

When sound transmits in a medium from the sound source (e.g., a speaker) to a receiver (e.g., human ear), it travels along a multitude propagation paths (see Figure \ref{fig16}), leading to sound pressure changes due to scattering, reflection, refraction, diffraction, etc. Sound propagation techniques aim to model this process. 

Realistic simulation of sound propagation greatly enhances the immersion of a scene by predicting the response in the environment for a given source signal and increasing the sense of presence. Sound and light share many properties, but they are also quite different. For example, the receiving frequency range of the sound wave and the light wave are different; because the sound propagation speed is far less than that of light, the time factor must be considered~\cite{Cha14}. Some of the earlier methods used for generating acoustic effects and sound rendering are based on reverberation filters~\cite{Par12}. In practice, these are simple and low-cost algorithms, which simulate the decay of sound in rooms. These filters are designed based on different parameters and are either specified by an artist or computed using scene characteristics~\cite{tsingos2009precomputing}. They can also be extended to compute directionally-varying reverberation effects~\cite{Ant13}. Many other precomputation methods for multi-player games or distributed 
virtual worlds are based on precomputed acoustic similarity metrics~\cite{taylor2014rendering}. Many of these methods have low runtime overhead  and can handle dynamic scenes. However, assume that the reverberant sound field is diffuse, making them unable to generate accurate directional reverberation or time-varying effects for general scenes.
Recent work in sound propagation can be classified into three categories, wave-based  methods, geometric methods, and hybrid combinations (see Table \ref{table2}).

\subsection{Wave-based methods}

Based on different divisions of the computational space, wave-based sound propagation methods can be classified into volume-based methods and surface-based methods. The former methods discretize the whole space into grid cells and perform a numerical calculation on each grid cell. The memory used is linear with the volume of the scene, which is suitable for scenes with high, specific surface areas (area/volume) such as indoor scenes~\cite{Meh13}. The latter methods transforms the surface of the scene into discrete elements and calculate for each element, which is suitable for scenes with low, specific surface areas. In a large scene or open outdoor area, the geometric objects are sparsely distributed and the sound wave propagation has more uniformity~\cite{Cheng05}\cite{Pind19}, so the surface-based wave-based methods are more suitable.

Among the volume-based wave-based methods, the most common are the finite element methods~\cite{Zie13}\cite{Tho06} and the time-domain finite difference methods~\cite{Yee66}\cite{Taf05}\cite{Sak06}. The finite element method first establishes the finite element model of the sound field then solves the corresponding finite element equation, thus obtaining sound field values for one or many sources and analyzing the sound field characteristics on this basis. Some existing finite element methods such as pseudospectral time domain (PSTD)~\cite{For88}\cite{Horn10} have achieved higher accuracy using a rough spatial discretization method. However, the computational cost of these methods is highly dependent on the size of the scene and the maximum simulation frequency of the acoustic signal. The computational cost is linear in scene volume and fourth power in max frequency. If we simulate sound propagation, doing a full bandwidth simulation on large scenes such as a concert hall, within a desktop computer's resources remains out of reach. Bilbao et al.~\cite{Bil09}\cite{Bil12}\cite{Bil15}\cite{Bil17}\cite{Bil19} used the finite difference methods for sound synthesis with physical models of musical instruments as well as room acoustics. In the early 1990's, Botteldooren~\cite{Bottel94}\cite{Bot95} and Savioja et al.~\cite{Savioja94} used finite difference time domain (FDTD) for computing room acoustics. Later, FDTD was widely used for sound field simulation. However, such methods assume that the difference quotient instead of the difference is used to solve the wave equation, and the impulse response (IR) is calculated. The FDTD method has a high accuracy, which can be applied to complex acoustic phenomena and complex obstacle scenes ~\cite{Osh13}\cite{Hei06}\cite{deG05}\cite{Dra13}. Raghuvanshi et al.~\cite{Rag08}\cite{Rag09}\cite{Rag10} proposed a fast method based on adaptive rectangular decomposition. This method solved the wave equation in rectangular domains to achieve high accuracy, even on grids approaching the Nyquist limit. The wave equation has analytical solutions in rectangular domains that can be efficiently computed with the Discrete Cosine Transform (DST). As a result, this method achieves at least an order of magnitude performance gain, both in terms of memory and computation, compared to standard FDTD implementation. Mehra et al.~\cite{Meh12} provided a parallel GPU implementation of the ARD technique which results in a further order-of-magnitude acceleration. Savioja~\cite{Sav18} proposed a real-time FDTD method suitable for low-frequencies and medium-frequencies based on GPU acceleration. In order to address FDTD's difficulties in simulating the reflection of sound waves on the impedance ground, i.e., the ground as an impedance boundary condition, Cott\'{e} and Blanc-Benon~\cite{Cott09} proposed obtaining the impedance time-domain boundary conditions by solving the linearized Euler equation. This method can simulate long-distance sound propagation on the impedance ground. Recently, Rosen et al. \cite{Rosen00} enabled real-time wave-based simulation on small fully dynamic scenes on a single CPU core by restricting simulation to a 2D slice of the scene at listener height.

The most common surface-based wave-based methods are the boundary element method~\cite{Fai03} and the equivalent source method ~\cite{Kro95}\cite{Och99}\cite{Pav06}\cite{Liu09}\cite{Meh13}. The boundary element method (BEM) transforms the three-dimensional domain problem into a two-dimensional boundary problem, which greatly reduces the computational complexity. Gumerov and Ramani demonstrated good results by combining fast multipole method and the boundary element method~\cite{Gum09}. However, there are many monopole and dipole sound sources in the final solution, which take a large amount of storage space. Therefore, the fast multipole BEM is still an off-line algorithm.

\begin{figure}[!t]
\centering
\includegraphics[width=2.8in]{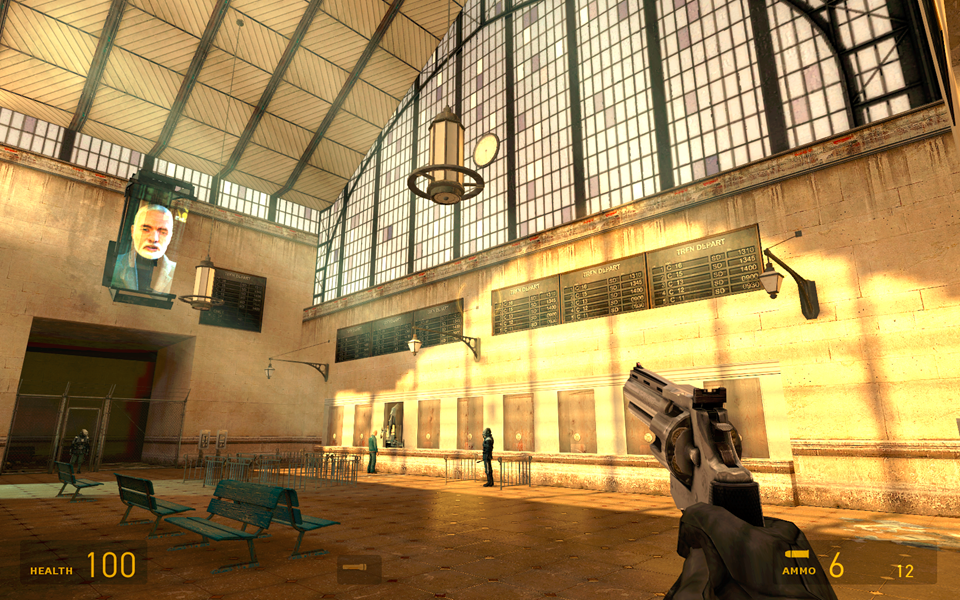}
\caption{The train station scene (http://source.valvesoftware.com) \cite{Rag10_2}.}
\label{fig17}
\end{figure}

Mehra et al.~\cite{Meh13} used the equivalent source method to solve sound propagation in large-scale open scenes. They assume the scene consists of disjointed objects and express their acoustic characteristics via a per-object transfer function that relates the amplitudes of the incoming equivalent sources to outgoing equivalent sources. Therefore, the acoustic response of a given scene to a static sound source can be formulated as solving a global linear system. This method can model realistic acoustic effects, including diffraction, focusing, scattering, and echoes, in a large-scale scene. This method needs to match the boundary conditions of the object surface by superimposing the sound field generated by multiple sources at multiple positions in the object. The number of equivalent sources limits the calculation efficiency, so it is not suitable for a scene with numerous sound sources and numerous radiator objects. However, memory is reduced by orders of magnitude compared to traditional methods and real-time sound propagation in a scene sized hundreds of meters can be realized.

Precomputation has generally been used to achieve real-time performance of sound propagation in static scenes, encoding the acoustic impulse response as a 7D function of sound sources and receiver position. However, this method suffers from a large amount of memory cost. To this end, Raghuvanshi et al.~\cite{Rag10_2} leveraged the impulse response-based technique to compress and extract acoustic information and realized interactive simulation of sound propagation in large-scale static scenes with dynamic sources based on pre-calculation. This method can achieve real-time sound propagation in the complex train station scene (Figure \ref{fig17}) with appealing effects, including diffraction and reverberation for dynamic objects and listeners. Later, Raghuvanshi and Snyder~\cite{Rag14} proposed a compactly parametric wave field coding method for precomputed sound propagation. This method encodes the wave function as a time-invariant 6D field based on four perceptual parameters, namely direct sound loudness (LDS), early reflection loudness (LER), early decay time (TER), and late reverberation time (TLR).

A sound field can provide a user with spatial and directional source information in a scene. Raghuvanshi and Snyder~\cite{Rag18} used the precomputed wave-based method to model the  directional effects. They designed a 9D directional response function related to listener position, time, and listener direction. Mehra et al.~\cite{Meh14,Meh15} designed a general framework to introduce source and listener directivity for wave-based sound propagation algorithms. Spherical harmonics (SH) are exploited for directional source expression by precomputation. They decomposed SH interactively and evaluated the sound field at the listener position by summing precomputed SH sound fields by weight. Chaitanya et al. \cite{Chaitanya2020} encoded and rendered the full 11-D bidirectional impulse response that has both source and listener directivity. Unlike ~\cite{Meh14}, CPU cost is insensitive to scene complexity and angular resolution of HRTF and source directivity function (SDF).

Ambient sounds such as wind, rain, and surf are quite common, and they serve as background sound in a scene. There are numerous sound sources in the scene, and the scene occlusion plays a critical role in indicating the sound directionality and spatial size. Modeling sound propagation in such scenes is challenging. Zhang et al.~\cite{Zhang18} assumed ambient sources consist of independent sound events (e.g., a raindrop) and approximated them using an ideal notion of spatio-temporal incoherence. They then developed a streaming encoder for directional distribution of the precomputed response via spherical harmonics for each listener position. 

\begin{figure}[!t]
\centering
\includegraphics[width=2.8in]{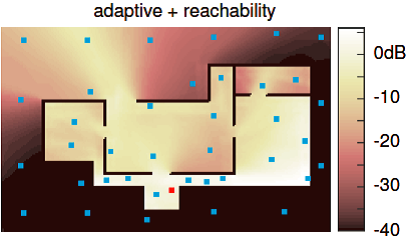}
\caption{Scene-aware sampling and interpolation \cite{Cha19}.}
\label{fig18}
\end{figure}

During runtime rendering, precomputed wave-based methods sample the impulse response at fixed probe positions to estimate the values at any dynamic listener position. Uniform sampling was usually adopted in prior works, which would make the computational overhead increase in narrow scenes. Chaitanya et al.~\cite{Cha19} proposed an adaptive sampling approach (see Figure \ref{fig18}) by introducing the “local diameter” measure for a given position. In order to generate smooth results, they also proposed a reachability-based interpolation technique for diffracted transport.

The above precomputed real-time methods all assume a uniform atmosphere with constant speed of sound. ESM cannot work without it, and neither can the ARD solver - both rely on the homogeneity assumption. So new methods are needed. Recently, Liu and Liu \cite{Liu20_2} proposed a sound propagation method in outdoor three-dimensional environments based on precomputation and sound signal encoding and decoding by considering the heterogeneity of outdoor atmospheres and constructed a sound propagation model accounting for ground effects and atmospheric heterogeneity. This method can realistically simulate outdoor sound propagation with high calculation efficiency and less memory consumption, meaning that it can achieve interactive simulation for a large-scale outdoor scene.

\subsection{Geometric propagation methods} 
Geometric methods assume that sound travels along a straight line, and determine the path of sound propagation according to energy attenuation. These methods are faster than wave-based methods and suitable for high-frequency sound propagation. These methods assume that the sound wavelength is far smaller than the obstacles in the scene. As a result it is difficult to accurately simulate some low-frequency acoustic phenomena such as diffraction. 

\subsubsection{Main Approaches}
Geometric propagation methods have been widely studied for more than five decades. 
One of the earlier and landmark papers in geometric sound propagation was presented by Krokstad et al.~\cite{krokstad1968}, which used ray tracing for computing time-energy responses and demonstrated the benefits to practical room acoustic design. 
Furthermore, some of the earlier and widely used methods to compute specular reflections are based on image sources~\cite{allen1979,borish1984}.  These techniques compute virtual sources from a given sound sources in a recursive manner for every reflection. The underlying complexity of the basic method can be exponential in the number of reflections. These methods can guarantee all specular paths up to a given order.

These earlier techniques based on ray tracing and image sources have been considerably extended and widely used. These include many  improved and faster methods for ray tracing that can also perform diffuse reflections~\cite{vorlander1989,dalenback1996room,Mic12,Schi14,schissler2018interactive}. In many ways, these ray tracing methods are the most popular techniques for geometric sound propagation and also integrated in current gaming and VR systems. The image source methods are also used frequently and their performance can be considerably improved using  beam tracing~\cite{dadoun1985geometry,stephenson1995pyramidal,Fun98,laine2009accelerated} and frustum tracing~\cite{Lau07,Cha09}.

One of the most general model for geometric room acoustic can be written as an integral equation. One of the first equations was the {\em Kuttruff's integral equation} for diffuse reflections in a convex room~\cite{kuttruff2016room}. Many extensions of this mathematical model have been proposed subsequently. Lewers~\cite{Lew93} proposed a model for room acoustics by combining beam tracing and radiant exchange. The room acoustic rendering equation~\cite{siltanen2007room} provides a framework that can be used to model most of the geometric acoustic methods for room acoustics. Many precomputation techniques from computer graphics have been extended to audio simulation and rendering based on the room acoustic rendering equation~\cite{antani2011direct,antani2012interactive}.

\begin{figure}[!t]
\centering
\includegraphics[width=3.0in]{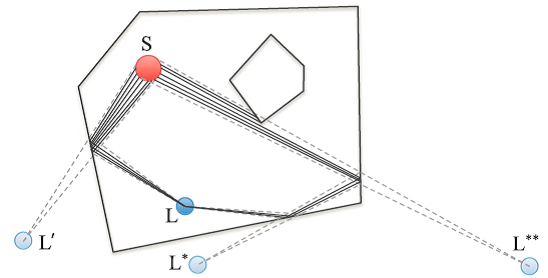}
\caption{Illustration of calculating specular sound for spherical sources \cite{Schi16}.}
\label{fig19}
\end{figure}

\subsubsection{Interactive Geometric Propagation}

Schissler et al.~\cite{Schi16} proposed a fast backward propagation method for multi-source scenes that estimated higher-order reflections from spherical sound sources (see Figure~\ref{fig19}). Experiments showed that this method achieves 5x speedup over traditional forward ray tracing methods. Antani et al.~\cite{antani2011direct} proposed a direct-to-indirect acoustic transfer operator to simulate diffuse reflection by introducing the idea of precomputed light transport algorithms. 

Antani et al.~\cite{antani2012interactive} proposed an interactive sound propagation algorithm that considers high-order reflections and edge diffractions in a complex scene with dynamic sound sources and a moving listener. The key is a precomputed acoustic transfer operator based on the Karhunen-Loeve transform. They combined a ray tracing algorithm and acoustic radiance transfer for efficient computation of early reflections and late reverberation. This method is fast with low memory cost. Schissler et al.~\cite{Schi14} achieved interactive reflections and high-order diffraction effects for large-scale scenes by combining radiosity and path tracing techniques to evaluate diffuse reflections in an iterative manner. In precomputation, a global edge visibility graph was designed for computing higher-order edge diffraction. A wavelength based simplification algorithm was also presented to greatly reduce the diffraction edge number for efficiency. This method can achieve plausible sound effects at interactive rates for large static and dynamic scenes with specular and diffuse reflections and higher order diffraction.

Geometry-based methods do not perform well for difficult-to-simulate, low-frequency sound effects, e.g., diffraction and occlusion. Rungta et al.~\cite{Run18} combined a precomputed diffraction kernel with the ray tracing method to simulate sound propagation at interactive rates. They proposed a source-placement approach for efficiency while computing the diffraction kernel of a given object. This method worked for highly tessellated and smooth objects and generated diffraction and occlusion effects that are closer to wave-based methods. 

In order to further reduce the calculation cost for ray tracing method, Heinz~\cite{Hei93} proposed a technique called diffuse rain, which connects the sound source or a listener with the points in the sound emission path. Its principle is similar to the shadow ray method in computer graphics that has been used for visual rendering. As a deterministic path tracing method, the acoustic beam tracing algorithm~\cite{Cam00} extended the acoustic line to a beam in a volume, which can detect the sound path in a larger range. The set of acoustic beam tracing methods include cone beam tracing, the triangle beam tracing, and the adaptive beam tracing. However, the beam tracing methods split the beam when it encounters geometric discontinuity. 

Ray tracing methods for indoor scenes usually assume constant speed of sound, whereas outdoor sound must assume a speed of sound that depends on pressure and temperature. Raspot et al.~\cite{Ras95} proposed a ray tracing method suitable for outdoor scenes, that gives a more real prediction, even in the low-frequency range. The geometric sound propagation model proposed by L'Esperance et al.~\cite{Les92}\cite{Les93} is based on the assumption that the ray path is a circle due to a vertically linear sound velocity profile, and the method can account for most of the sound energy attenuation mechanisms such as ground effect, atmospheric absorption, turbulence, and wind effects, however; this model cannot simulate sound propagation in the ``shadow area," i.e., the area that cannot be reached by a sound wave due to obstacles or the refraction effect. In order to overcome this shortcoming, Berry et al.~\cite{Ber88} proposed predicting the sound pressure in the shadow area formed in the upward refraction atmosphere. In order to handle sound propagation in complex outdoor terrains and atmospheric heterogeneous media, Lamancusa et al.~\cite{Lam93} proposed an advanced ray tracing model. They employ the axisymmetric ray path equation to generate the ray path and then calculate the propagation loss, absorption, and refraction degree of each ray as it hits on the ground. They then use the numerical grid division to integrate the changes of the shape of the complex terrain and acoustic velocity into the discrete path. Prospathopoulos et al.~\cite{Pro07} used the ray tracing method to determine the sound profile in the near earth propagation modeling, and more accurately simulated the near earth sound field. Mo et al.~\cite{Mo15} considered the influence of wind, temperature, and other environmental factors, combined the adaptive grid with parabolic ray tracing, and solved the problem of non-linear sound propagation. This method achieved significant performance improvements over prior methods but did not consider the simulation of low-frequency phenomenon of sound and the impact of ground effects on sound propagation. They combined the analytical ray curve tracker with the Gaussian beam model~\cite{Mo17} to calculate the sound field at near interactive speeds.

Funkhouser et al.~\cite{Fun98} used precomputed spatial subdivision and a beam tracing algorithm for real-time acoustic simulation in virtual environments.  This method can estimate reverberation at interactive rates and supports the model's high order reflections. Based on the fact that sound can only be heard at the positions of ``avatars," Funkhouser et al.~\cite{Fun99} proposed a real-time acoustic modeling method for distributed virtual environments (VDEs), efficiently computing reverberation paths by combining three beam tracing algorithms: the priority-driven beam tracing algorithm, the bidirectional beam tracing algorithm, and the amortized beam tracing algorithm. Most of the above geometric methods can model direct sound transmission and reflection while ignoring diffraction or they can simply approximate the diffraction effect.  

\begin{figure}[!t]
\centering
\includegraphics[width=3.0in]{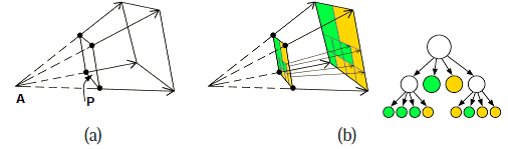}
\caption{Illustration of AD-Frustum. (a) is a frustum, represented by corner rays and apex $A$, (b) a hierarchical frustum \cite{Cha09}.}
\label{fig20}
\end{figure}

Lauterbach et al.~\cite{Lau07} presented a frustum tracing algorithm for sound propagation. They used a four-sided convex frustum for efficient intersection tests during hierarchy traversal, intersection, and reflection interactions. Different from beam tracing and pyramid tracing, this method performed clipping using ray packets. Owing to the volumetric formulation, this method can achieve interactive sound rendering in complex scenes with dynamic sources and objects. Chandak et al.~\cite{Cha09} proposed AD-Frustum, an adaptive frustum tracing method for interactive sound propagation. They used frusta for accurate sound propagation path generation, and each AD-Frustum consisting of an apex and a quadtree represents the subdivision of a frustum corresponding to a bundle of rays (see Figure \ref{fig20}). These methods were the first set of algorithms to geometric sound propagation at interactive rates on a common PC. 

\begin{figure*}[!t]
\centering
\includegraphics[width=6.0in]{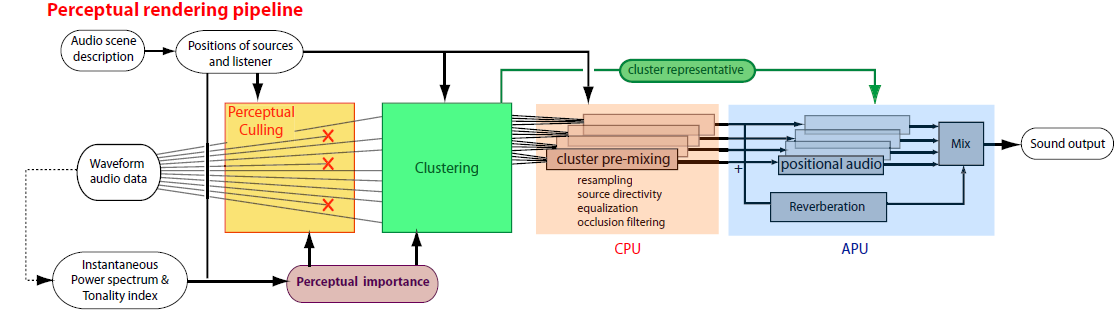}
\caption{The perceptual sound rendering pipeline for complex scenes with hundreds of moving sound sources \cite{Tsi04}.}
\label{fig21}
\end{figure*}

Cao et al.~\cite{Cao16} proposed a new bidirectional sound tracing algorithm, called Bidirectional Sound Transport (BST), which converges faster and balances the early reflection and the late reflection pulse response. This method can achieve interactive sound propagation effects for large-scale scenes with complex occlusions on a common PC. By introducing multiple rounds of importance sampling, this method works well for scenes containing multiple sound sources while achieving fine sound propagation effects, e.g., generating energy response with up to a 7 dB higher signal-to-noise ratio. Schissler and Manocha~\cite{Schi16_2} proposed an adaptive impulse response modeling approach for interactive sound propagation in scenes with multiple sound sources. They used impulse response cache and an adaptive ray tracing method that leverages psycho-acoustic properties of the IR length. Experiments showed this method traced fewer rays while preserving high fidelity for real-time sound propagation in complex scenes.

\subsection{Diffraction Effects}

Most of the above methods assume that sound travels in a homogeneous medium with a constant speed, which is difficult to extend to the sound propagation calculation in non-uniform media. In addition, due to the diffraction effect of sound waves, the geometrical acoustic technology needs to ensure that the wavelength is smaller than the object scale, which limits its application in the low-frequency domain.

Many techniques were proposed to model diffraction effects with geometric acoustic methods. The two primary diffraction models used in geometrical
acoustic simulations are based on the Uniform Theory of Diffraction (UTD)~\cite{kouyoumjian1974uniform} and the Biot-Tolstoy-Medwin (BTM)~\cite{biot1957,medwin1982} method. UTD is a less accurate frequency-domain model of diffraction for infinite edges that can generate plausible results for
interactive simulation in certain scenarios. It has been used for geometric sound propagation based on ray tracing~\cite{Schi14}, beam tracing~\cite{Tsi01}, and frustum tracing~\cite{taylor2009fast}. The UTD method has lower computational requirements and has been used for several interactive simulations. The BTM model is an accurate time-domain diffraction formulation that evaluates an integral of diffracted sound along finite rigid edges, can be extended to higher order diffraction, and can be combined with wave-based methods~\cite{roman2016hybrid,svensson1999,svensson2006use}. It can be accelerated using from-region visibility computations~\cite{antani2012efficient} or edge subdivisions~\cite{calamia2006fast}. The BTM method is considered more accurate than UTD and can be used with finite edges.

\nocite{antani2012interactive}

\subsection{Hybrid methods}
Hybrid sound propagation methods seek to strike a balance between performance and efficiency by coupling different techniques to achieve accurate and fast sound propagation simulation~\cite{Yeh13}. The existing hybrid methods are mainly based on the decomposition of the sound field space or frequency. Granier et al.~\cite{Grainer96} used FEM and BEM for low frequencies and an augmented geometrical acoustics model for high frequencies. One hybrid method based on frequency decomposition~\cite{Are12} divided the spectrum of the input sound signal into low-medium frequency and high frequency. The former uses accurate numerical acoustic method for modeling, while the latter is based on a geometric method that is good at handling high-frequency sound. Hybrid numeric-geometric algorithms have been proposed to approximate diffraction and occlusion effects around smooth surfaces~\cite{Run18,Tang20_1}. Other hybrid method is based on combining ray tracing with reverberation filters~\cite{schissler2018interactive}, which has very low runtime computational overhead and also works well on mobile devices.

\subsection{Psycho-acoustic evaluation}
Rungta et al.~\cite{Rung16} designed two experiments to analyze the psychoacoustic characterization of the sounding effects of diffraction and reverberation. They found that accurate sound propagation methods increase one’s perceptual differentiation ability. 
Rungta et al. \cite{Rung18} further presented a user study to evaluate the influence of reverberation and spatialization on word identification performance in a multi-talker environment. They indicated that spatialization significantly improves the target-word recognition performance, while reverberance has a negative impact, causing the loss of 1.5 dB on average. Recently, Rungta et al. \cite{Rung18_2} proposed a perceptual acoustic evaluation metric, P-Reverb, for early and late reflections.



\section{Sound rendering}

\begin{figure}[t]
\centering
\includegraphics[width=3.5in]{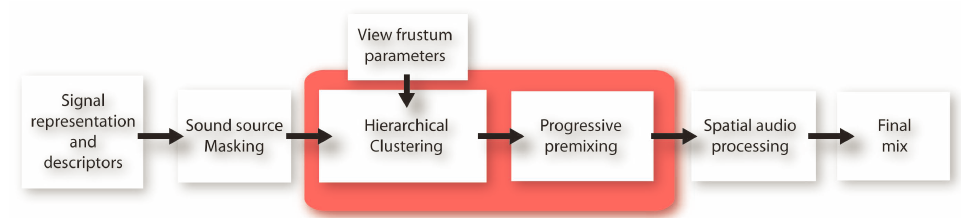}
\caption{The pipeline of progressive perceptual audio rendering for complex scenes \cite{Moe07}.}
\label{fig22}
\end{figure}

Based on an importance sampling strategy, Wand and Stra$\ss$er~\cite{Wan04} designed a stochastic sampling approach for an approximate rendering of scenes with a large number of sound sources. This approach includes two main threads, a sampling thread that selects representative sound sources and a mixing thread that blends the sampled sound sources. However, this method is efficient only in cases with static sound sources. Tsingos et al.~\cite{Tsi04} proposed dynamically eliminating inaudible sources before grouping the remaining audible sources and explored a real-time 3D audio rendering framework (see Figure \ref{fig21}) for scenes with moving sound sources. Moeck et al.~\cite{Moe07} developed a novel  progressive perceptual sound rendering pipeline (see Figure \ref{fig22}). This method can achieve high-quality rendering of thousands of sound sources on a "gamer-style" PC by introducing an improved hierarchical sound source clustering strategy and a premixing technique for progressive per-source processing.

Schissler and Manocha~\cite{Schi18} proposed a novel sound rendering pipeline by combining ray-tracing based methods with reverberation filters. They used the spherical harmonic basis functions to represent the direct sound, early reflections, ad late reverberation, which can capture details of the impulse response. This method can be used on mobile devices, as its computational requirements are low. Chaitanya et al. \cite{Chaitanya2020} proposed fast multi-source reverb rendering using the "canonical filter" approach \cite{Rag14}\cite{Rag18}. The key property of the canonical filters approach is that it is able to approximately render per-source reverberation characteristics without incurring the cost of instantiating a reverb filter per source. The trick is to design a set of fixed "canonical" filters that have unit initial energy, while varying their RT60 and direction at listener. At runtime, any dynamic per-source perceptual acoustic parameters can be mapped to linear weights for these filters. The per source cost is thus reduced to a scaled sum of the source signal at the input to these fixed filters.

The Feedback Delay Network (FDN) approach was proposed in a pioneering work by Jot and Chaigne \cite{Jot1991}. This method provided a general design structure for reverberators that can separately and independently control the energy storage, damping and diffusion components of the reverberator. However,  for games and VR, if there are 100 sources, it can not afford to instantiate 100 reverb filters, even fast FDNs are too expensive for that. Therefore, one needs algorithms with (ideally) sub-linear scaling in sound source count, usually by exploiting perception \cite{Tsi04}. Schissler and Manocha \cite{Schi16} developed an interactive sound propagation and rendering for large multi-source scenes by leveraging source-culling.

Speakers and immersive head-mounted displays (HMDs) are two widely used devices for sound display. Researchers studied multi-channel loudspeaker arrays and the WFS and Higher-order Ambisonics format or discrete-channel formats that are common in games. In HMDs, the personalized head-related transfer functions (HRTFs) are  utilized to perform accurate simulation and realistic display of spatial audio filtered by the human body and head. HRTFs are different for each individual, because they are related to the shapes of the outer ears (pinnae), head, and torso. Some research \cite{Bil14}\cite{Mes14}\cite{Zot04}\cite{Mes14_2} has been conducted to compute the person-aware HRTFs for accurate sound display. 

Savioja et al.~\cite{Savioja99} presented a complete real-time system for room acoustic auralization with an approach of detailed early reflections and simplified reverberation. Another related topic is visualization of acoustic measurements. Stettner and Greenberg~\cite{Ste89} used both computer simulation and visualization techniques to help describe and evaluate the acoustical behavior of performance halls. 

In modern game engines, game audio middleware such as Wwise or FMOD $\footnote{http://www.wwise.com; https://fmod.com}$ are used to add sound or music to any game. The resulting API receives events from the game engine and  provides tightly coupled, yet independent workflow between visuals and audio.

\section{Deep sound simulation} 

\subsection{Deep audio synthesis for video}
Video acquisition is becoming more and more convenient in daily life. Given a video, it still remains challenging, however, to add realistic audio that is synchronized with the animation in the video. Li et al. \cite{Li18} proposed adding scene-aware ambisonic audio to $360^o$ videos for indoor scenes. Given specified sound sources, this method first recorded the acoustic impulse response with a simple microphone and a speaker. They used a geometric-based method to simulate of the early reverberation, which was then improved with a frequency modulation method. The late reverberation was separated from the recording. This method is efficient and can produce audio synchronized with the $360^o$ video, providing a practical method for audio post-production.

Adding audio for a given video can also be viewed as a cross-modal transformation task. The difficulty of video-to-audio transformation lies in the fact that video and audio belong to two different types of signals, which are quite different from each other in terms of expression and features. It is difficult to find the correlation between them artificially. Fortunately, popular modern deep learning techniques provide a novel means to tackle this issue by directly learning the mapping between vision and sound. 

Owens et al.~\cite{Owe16} proposed a network framework based on CNN and LSTM to realize automatic percussion synthesis. Chen et al.~\cite{Chen17} designed a network tailored for musical instrument sound synthesis based on GAN. The two sound synthesis algorithms discussed above are based on the cochlear diagram and the mel-spectrum, respectively. Therefore, the original audio signal cannot be directly generated. Because audio and video have different time-space scales and mismatched feature structures, it is more direct and challenging to build an end-to-end cross-modal environment sound synthesis framework. By training videos collected from outdoor environments, Zhou et al.~\cite{Zhou18} synthesized a more realistic environment sound based on the sampleRNN model. However, experiments by Owens and Mehri et al.~\cite{Owe16}\cite{Meh17} showed that RNN has limitations in terms of visual content learning, so that this method cannot achieve an ideal synchronization effect.

Cheng et al.~\cite{Cheng19_3} proposed an end-to-end cross-modal ambient sound synthesis method by combining CGAN and SampleRNN to improve the scalability and synchronization of existing deep learning-based methods. The video depth feature is extracted based on the VGG network model. Then, through a novel timing synchronization network model, the cross-modal feature transformation from video to audio with higher synchronization was realized.
Audio-visual neural networks have also been designed to classify objects~\cite{Sterling_2018_ECCV}, to pour liquids using a robot~\cite{8968118}, and to track an object~\cite{9197528}.

\subsection{Deep learning for sound synthesis}
Jin et al.~\cite{Jin20} proposed Deep-Modal for impact sound synthesis. They encoded the mode data as fixed-length vectors and developed a network with an encoder-decoder architecture to map shape features of 3D objects to mode data. This method can produce interactive sounds of objects with various shapes. Hawley et al. \cite{Hawley2020} and Ji et al. \cite{Ji2020} provided reviews on the deep learning methods for musical instrument sound synthesis.

\subsection{Deep learning for sound propagation}

Pulkki and Svensson~\cite{Pul2019} proposed a machine learning method for estimating filter-parameters from the object geometry and they rendered the scattering effect with a parametric filter structure. Fan et al.~\cite{Fan20} evaluated sound propagation effects such as scattering and occlusion from objects by using CNN. They mapped the scattering cross section due to a convex scatter to a 2D image with spatial loudness distribution. Thus, they formulated the sound propagation as an image-to-image regression problem. Tang et al.~\cite{Tang20_1} proposed learning the acoustic scattering field of objects for lower frequencies from 3D geometric object information. Then, at runtime, the above learned information can be used for estimating the scattered field when sound waves interact with objects. This approach can handle objects undergoing deformation. 

\begin{figure}[!t]
\centering
\includegraphics[width=3.5in]{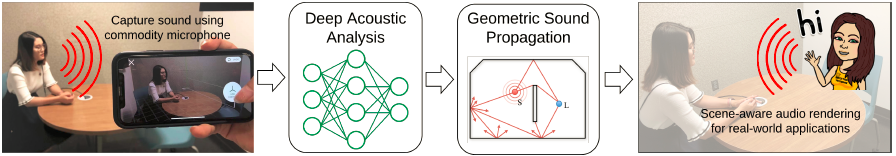}
\caption{The framework of the scene-aware audio rendering via deep acoustic analysis \cite{Tang20}.}
\label{fig15}
\end{figure}

\subsection{Sound rendering in Real-World Scenes}
Schissler et al.~\cite{Schi17} proposed producing acoustic effects in captured real-world scenes via CNN for multimodal augmented reality. They used CNN to derive the sound material properties, e.g., absorption coefficients, which were then used to simulate the effects of sound propagation. This method works well for indoor scenes.

Tang et al.~\cite{Tang20} used deep learning for simulating sound propagation effects for real-world scenes and proposed a scene-aware audio rendering method (see Figure \ref{fig15} for the framework). They recorded the sound signals in a real-world room using a cellphone microphone. The acoustic material properties and the frequency equalization of the room were computed by a deep learning network, and these parameters were then used to estimate sound effects in a virtual model of the room. 

Recently,  Morgado et al.~\cite{Morgado18} presented a method to convert an mono audio recorded by a $360^o$ video camera
into spatial audio, that is a novel representation form of sound over a sphere. Huang et al.~\cite{Huang19} proposed audible panorama, that augments panorama images through realistic audio assignment.

\section{Conclusion and future directions}
Research on sound simulation has made great progress in recent years. Many methods have been developed that have achieved excellent results in both sound quality and computational efficiency, and some of these methods have even been applied in computer games, virtual reality, film foley, etc. Many game engines and VR applications have incorporated geometric and wave-based sound propagation technologies, including  Microsoft Project Acoustics~\cite{MicrosoftAcoustics}, Oculus Spatializer~\cite{OculusS}, and Steam Audio~\cite{SteamA}. However, there are still challenges that must be addressed.

(1) Currently, almost all the videos generated using computer graphics or visual rendering systems are sort of silent. There will large amounts of work to do to make graphics content "audible" which is very challenging and exciting.

(2) Although some types of sounds can be synthesized, there are still many other, more complex sounds that have not been explored. Since it is easy to capture videos in daily life, a future direction may be to realize sound synthesis by computing sounds from videos, such as adding scene-aware ambisonic audio to $360^o$ videos for indoor scenes \cite{Li18}. Moreover, existing sound synthesis and sound propagation methods usually suffer from problems of large computational overhead, large memory, and trivial parameters tweaking. It is necessary to develop lightweight sound simulation models in the future. 

(3) It seems promising to apply modern deep learning techniques to sound simulation research. A few attempts have been made in this direction, such as CNN \cite{Tang20}, CGAN \cite{Cheng19_3}, and V2RA-GAN \cite{Liu21}, which have been used for sound synthesis; CNN was also exploited to derive the sound material properties to replace tedious manual tweaking. Acoustic scattering can also be regarded as an image-to-image regression problem \cite{Fan20} in which a CNN was trained to learn the mapping of simple geometries such as convex prisms to the acoustic loudness field.

(4) Prior studies separated sound synthesis from sound propagation and simulated each of them independently. It may be beneficial to couple sound synthesis and sound propagation as a whole and build an entire process of sound simulation, such as SynCoPation \cite{Run16}. SynCoPation used a single-point multipole expansion (SPME) to merge sound radiation and propagation for every surface vibration mode of a given rigid object, ray tracing was adapted for computing the impulse response of each mode in sound propagation, and a perception-aware Hankel function was exploited to reduce the computational cost of this acoustic propagation procedure.
 
(5) There is still room for improving the efficiency of current sound propagation methods. It is challenging to handle sound propagation in large-scale, outdoor, dynamic scenes, especially in terms of generate low-frequency effects. The most accurate methods are based on wave-based propagation, but they are mostly limited to static scenes, though learning methods are promising and can approximate the scattering field of moving or deforming objects~\cite{Meng21}.

(6) It is necessary to develop plausible, quantitative sound evaluation metrics in the future. Current sound synthesis and sound propagation methods can produce more and more accurate results. However, it remains a challenge to access whether or not a sounding result improves the existing algorithms in terms of the perceptual sound quality.

(7) Sound simulation techniques can also benefit acoustic design \cite{Monks00} and musical instrument design \cite{valimaki96}\cite{valimaki06}\cite{Bha15}\cite{Ren15}. It will be a promising direction to combine advanced sound simulation and sound propagation techniques with the latest 3D printing technology to develop computer aided musical instrument (or room acoustics, etc.) design systems.

(8) Sound simulation methods are combined with machine learning and optimization methods \cite{Zhang2021} and are used for analyzing or processing real-world audio signals. These techniques are used in various applications including synthetic dataset generation for automatic speech recognition\cite{tang2019regression} \cite{tang2019improving} \cite{tang2019lowfreq} \cite{tang2021sceneaware} \cite{Ratnarajah21} \cite{ratnarajah2021irgan} and sound source separation~\cite{jenrungrot2020cone}, speaker placement~\cite{morales2019receiver}, acoustic material design optimization~\cite{morales2016efficient}, 3D reconstruction~\cite{zhang20173d, Ste19}, sound source localization~\cite{an2018reflection, an2019diffraction}, audio distractors for redirected-walking in a virtual environment~\cite{Rewkowski2019}, etc. There are plenty of challenges in terms of far-field automatic speech recognition, source separation, localization in noisy environments, etc.

\section{Acknowledgements}
This work was supported in part by the National Natural Science Foundation of China under grant nos. 62072328 and 61672375, ARO grant W911NF-18-1-0313, NSF grant no. 1910940, Adobe Research, and Intel. We are grateful to Lauri Savioja, Peter Svensson, Nicolas Tsingos, Changxi Zheng, and Nikunj Raghuvanshi for their feedback.


%



\ifCLASSOPTIONcaptionsoff
  \newpage
\fi



%

%

\begin{IEEEbiography}[{\includegraphics[width=1in,height=1.25in,clip,keepaspectratio]{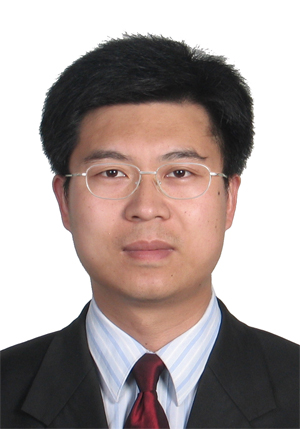}}]{Shiguang Liu} (Senior Member, IEEE)
received the PhD from the State Key Lab of CAD \& CG, Zhejiang University, P.R. China.
He is currently a professor with School of Computer Science and Technology, College of Intelligence and Computing, Tianjin University, P.R. China. His research interests include computer graphics, image/video editing, visualization, and virtual reality, etc. 
\end{IEEEbiography}

\begin{IEEEbiography}[{\includegraphics[width=1in,height=1.25in,clip,keepaspectratio]{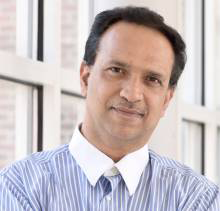}}]{Dinesh Manocha} (Fellow, IEEE)
is Paul Chrisman-Iribe Chair in Computer Science \& Electrical and Computer Engineering and Distinguished University Professor at University of Maryland College Park. He is also the Phi Delta Theta/Matthew Mason Distinguished Professor Emeritus of Computer Science at the University of North Carolina at Chapel Hill. His research interests include virtual environments, physically-based modeling, and robotics. His group has developed a number of packages for multi-agent simulation, robot planning, and physics-based modeling that are standard in the field and licensed to more than 60 commercial vendors. He has published more than 600 papers \& supervised 40 PhD dissertations. He is an inventor of 10 patents, which are licensed to industry.  He is a Fellow of AAAI, AAAS, ACM, and IEEE, member of ACM SIGGRAPH Academy, and Bézier Award from Solid Modeling Association. He received the Distinguished Alumni Award from IIT Delhi the Distinguished Career in Computer Science Award from Washington Academy of Sciences.  He was a co-founder of Impulsonic, a developer of physics-based audio simulation technologies, which was acquired by Valve Inc in November 2016. See http://www.cs.umd.edu/~dm
\end{IEEEbiography}







\end{document}